# Angular anisotropy in pre-fission neutron spectra and PFNS of $^{240}$Pu(n, F)

V. M. Maslov[1]

*220025 Minsk, Byelorussia*

Angular anisotropy of secondary neutrons was evidenced in neutron emission spectra (NES) of $^{239}$Pu+$n$ in 1972, and in prompt fission neutron spectra (PFNS) of $^{239}$Pu($n$, $F$) in 2019, it might be predicted for $^{240}$Pu($n$, $F$) PFNS now. In case of NES angular anisotropy is due to direct excitation of collective levels and pre-equilibrium/semi-direct (states in the continuum are excited) mechanism of neutron emission of first neutron in ($n$, $nX$)$^1$ reaction, in case of PFNS it is due to ($n$, $xnf$)$^{1,\ldots x}$ exclusive spectra of pre-fission neutrons. In $^{239}$Pu($n$, $xnf$) and $^{240}$Pu($n$, $xnf$) reactions observed PFNS envision different response to the emission of ($n$, $nf$)$^1$ neutron in forward or backward direction. Energies of ($n$, $nf$)$^1$ neutrons and their average values $\langle E_{nnf}(\theta)\rangle$ and other observables depend on the angle of emission $\theta$ with respect to the incident neutron momentum. Exclusive spectra of ($n$, $xnf$)$^{1,..x}$ neutrons at $\theta$~90$^o$ are consistent with $^{240}$Pu($n$, $F$)( $^{239}$Pu($n$, $F$), $^{239}$Pu($n$, 2$n$)) cross sections and NES of $^{239}$Pu+$n$ at $E_n$≲20 MeV. The correlations of the angular anisotropy of PFNS with the contribution of the ($n$, $nf$) fission chance to the fission cross section and angular anisotropy of pre-fission neutron emission are ascertained. The exclusive spectra of $^{240}$Pu($n$, $xnf$)$^{1,..x}$, $^{240}$Pu ($n$, $xn$)$^{1,..x}$ and $^{240}$Pu($n$, $n\gamma$) reactions are calculated with Hauser-Feshbach formalism as $^{240}$Pu($n$, $F$) and $^{240}$Pu($n$, $xn$) cross sections, angular anisotropy of ($n$, $nX$)$^1$ neutron emission being included. The influence of forward and backward emission of $^{240}$Pu($n$, $xnf$)$^{1,..x}$ pre-fission neutrons on PFNS are predicted to be stronger than observed for PFNS of $^{239}$Pu($n$, $F$). The ratios of $^{240}$Pu($n$, $F$) PFNS average energies $\langle E\rangle$ for forward and backward emission of $^{240}$Pu($n$, $xnf$)$^{1,..x}$ pre-fission neutrons are predicted to be similar to those observed for $^{239}$Pu($n$, $F$) PFNS. PFNS $\langle E\rangle$ are consistent with the measured data up to the threshold of $^{240}$Pu($n$, 2$nf$) reaction.

## I. INTRODUCTION

The prompt fission neutron spectra (PFNS) of fissile [1–4] and fertile [5–8] target nuclides neutron-induced fission seem to be different from each other a lot. Calculated prompt fission neutron spectra of $^{240}$Pu($n$, $F$) [6, 9, 10], should look much similar to those of $^{238}$U($n$, $F$) [5–7] and $^{232}$Th($n$, $F$) [5, 11], however, might be also different in a number of respects. Detailed measurement of observed prompt fission neutron spectra PFNS of $^{240}$Pu($n$, $F$) just appeared in [12]. The newest [12] and preliminary data for $\varepsilon$~0.89–10 MeV [13–15] on PFNS of $^{240}$Pu($n$, $F$) are almost consistent with the calculated estimates of average energies $\langle E\rangle$ and PFNS shapes [6, 9, 10], though with some exceptions. The newest angle-integrated data [12] provide PFNS for prompt fission neutron energy range $\varepsilon$~0.8–10 MeV, the extrapolations to outside energies need robust PFNS theory or reliable semi-phenomenology.

The reliability of the modelling of PFNS for neutron-induced fission of $^{239}$Pu might be augmented by comparing $^{239}$Pu($n_{\rm th}$, $f$) and $^{240}$Pu($sf$) [16] data sets. Similar augmentation is possible for PFNS of a pair of fission reactions, $^{241}$Pu($n_{\rm th}$, $f$) and $^{242}$Pu($sf$) [16]. In reaction $^{239}$Pu($n_{th}$, $f$) the main part of neutrons yield from $J^\pi = 0^+$ states, the same as in $^{240}$Pu($sf$) spontaneous fission neutron spectra (SFNS). Comparison of PFNS for $^{239}$Pu($n_{th}$, $f$) and SFNS of $^{240}$Pu($sf$) [7] in [1, 6] shows that at ε<0.2 MeV PFNS and SFNS of fissioning nuclide $^{240}$Pu depend only weakly on the excitation energy $U$. At energies ε≳$\langle E\rangle$ the PFNS of $^{239}$Pu($n_{th}$, $f$) is much harder than SFNS of $^{240}$Pu($sf$). The same would happen in case of calculated $^{241}$Pu($n_{\rm th}$, $f$) PFNS and measured SFNS of $^{242}$Pu($sf$). For a pair of PFNS of the reactions $^{240}$Pu($n_{\rm th}$, $f$)&$^{241}$Pu($sf$) such augmentation is hardly would be ever possible, though some guidance stems from comparing a pair of spectra of $^{240}$Pu($n_{\rm th}$, $f$) and $^{240}$Pu($sf$) [6]. Some guidance comes also from comparison of calculated $^{240}$Pu($n_{\rm th}$, $f$) PFNS and $^{240}$Pu($n$, $f$) PFNS data [12–15] at $E_n$~1–2 MeV.

---

[1]mvm2386@yandex.ru

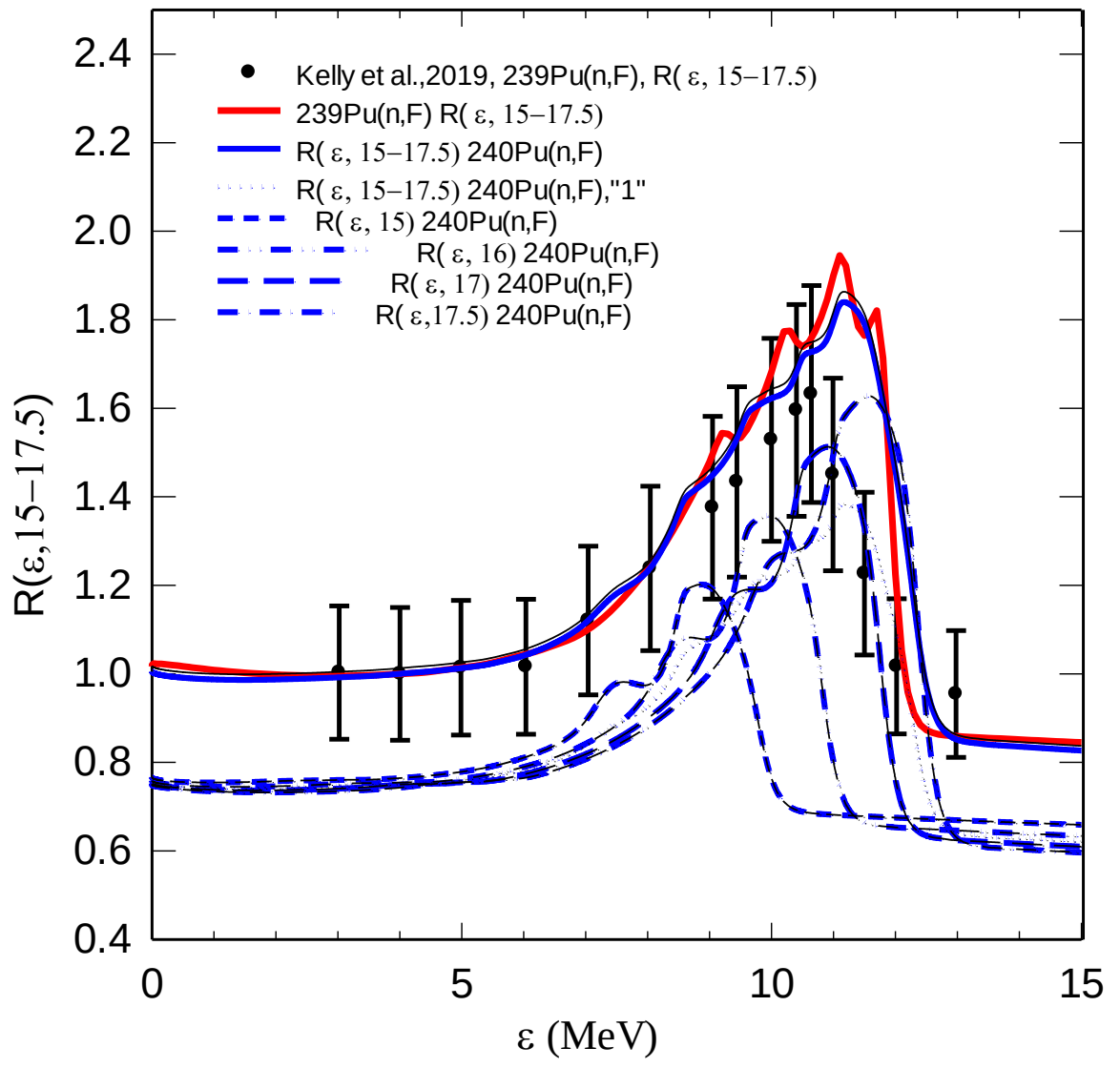

Fig. 1 Ratio $R^{\exp} = S(\varepsilon, E_n \approx 15-17.5, \Delta\theta) / S(\varepsilon, E_n \approx 15-17.5, \Delta\theta^1)$ for $^{239}$Pu($n$, $F$) PFNS and calculated ratio $R(\varepsilon, E_n, \Delta\theta, \Delta\theta^1)$ for "forward" ( $\Delta\theta$ ~35º–40º) and "backward" ( $\Delta\theta^1$=130º–140º) neutron emission. Data points: ● – $^{239}$Pu($n$, $F$) [17]; full blue line – $^{240}$Pu ($n$, $F$) PFNS, equated at $\varepsilon$~3–5 MeV; red dash-dotted line – $^{239}$Pu($n$, $F$) PFNS, equated at $\varepsilon$~3–5 MeV; partials of $R(\varepsilon, E_n, \Delta\theta, \Delta\theta^1)$ at $E_n$ ~15-17.5 MeV for $^{240}$Pu ($n$, $F$), normalized to unity; $E_n$ ~ 15 MeV – dotted blue line; $E_n$ ~ 16 MeV – short dashed blue line. $E_n$ ~ 17 MeV – double dotted dashed blue line; $E_n$ ~ 17.5 MeV – long dashed blue line.

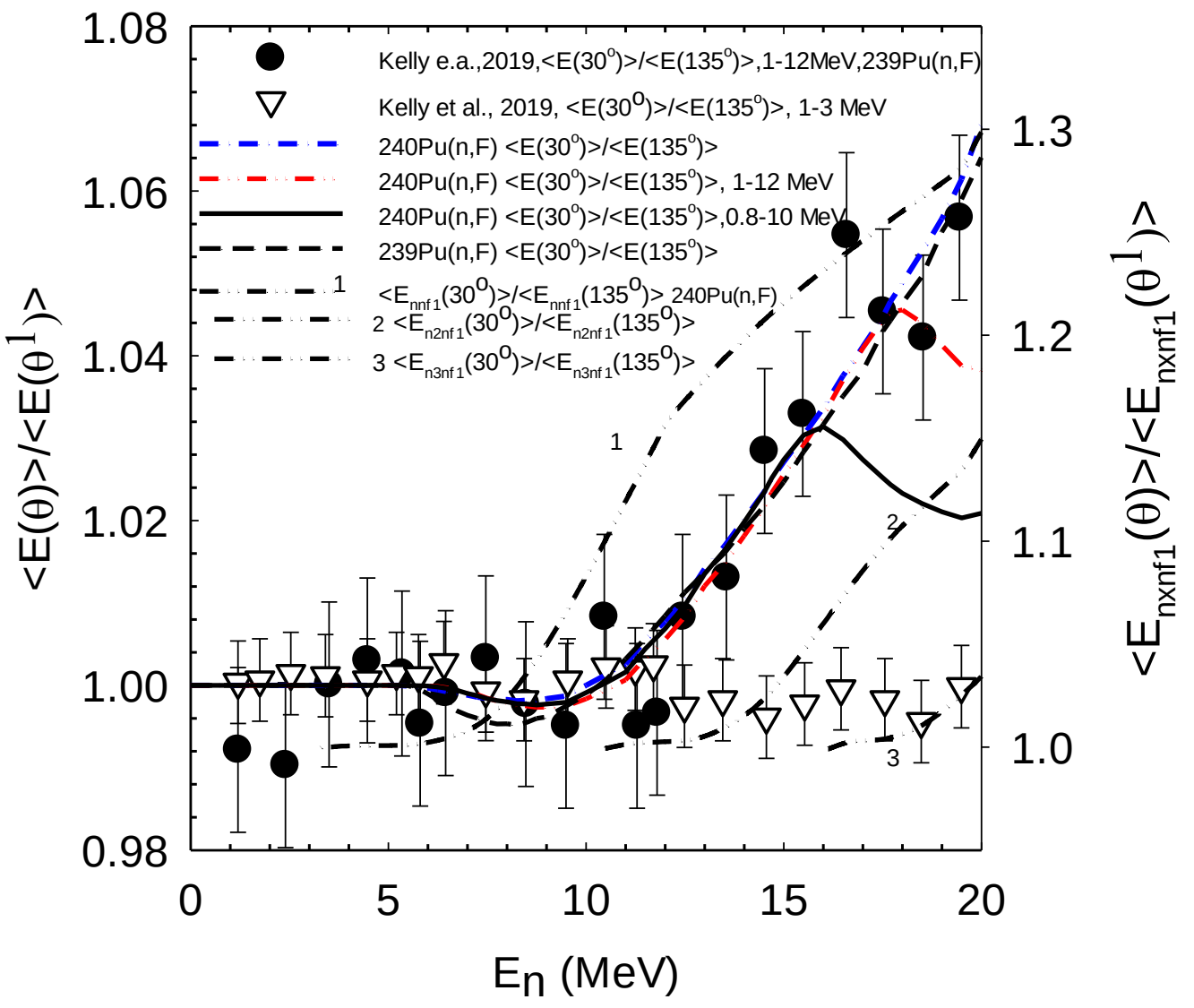

Fig. 2. PFNS of $^{239}$Pu($n$, $F$) and $^{240}$Pu($n$, $F$), ratio $\langle E(\theta)\rangle / \langle E(\theta^1)\rangle$. Data points: ▲ – $^{239}$Pu($n$, $F$) $\langle E(\theta \approx 30^o)\rangle / \langle E(\theta^1 \approx 135^o)\rangle$, $\varepsilon$~1–12 MeV [17]; ∇ – $^{239}$Pu($n$, $F$), $\langle E(\theta \approx 30^o)\rangle / \langle E(\theta^1 \approx 135^o)\rangle$, $\varepsilon$~1–3 MeV [17]. Full line – $^{240}$Pu($n$, $F$) $\langle E(\theta \approx 30^o)\rangle / \langle E(\theta^1 \approx 135^o)\rangle$, $\varepsilon$~0.8–10 MeV; blue dash-dotted line – $^{240}$Pu($n$, $F$), $\langle E(\theta \approx 30^o)\rangle / \langle E(\theta^1 \approx 135^o)\rangle$, $\varepsilon$~0–20 MeV; red dash-double dotted line – $^{240}$Pu($n$, $F$) $\langle E(\theta \approx 30^o)\rangle / \langle E(\theta^1 \approx 135^o)\rangle$, $\varepsilon$ ~1–12 MeV; dashed line –$^{239}$Pu($n$, $F$), $\langle E(60^o) / E(90^o)\rangle$, $\varepsilon$~0–20 MeV; lines 1, 2, 3 –$^{240}$Pu($n$, $F$) , $\langle E_{n,xnf}(\theta \approx 30^o)\rangle / \langle E_{n,xnf}(\theta^1 \approx 135^o)\rangle$, $x$ = 1, 2, 3.

When the PFNS of $^{240}$Pu($n_{\rm th}$, $f$) reaction is eventually fixed, the modelling of PFNS, as described in [1, 6], produce acceptable fits of available data on $\langle E\rangle$ and measured $^{240}$Pu($n$, $f$) and $^{240}$Pu($n$, $F$) PFNS data [12–15] at $E_n$~$\langle E \rangle$, $E_n$~$E_{nnf}$ and $E_n$~$E_{n2nf}$, here $E_{nxnf}$ is the threshold of (*n, xnf*) reaction. Major parameters of $^{240}$Pu($n_{th}$, *f*) PFNS modelling define the kinetic energy of the fragments at the moment of prompt fission neutron emission from the fragments. That may influence the $\langle E \rangle$, however, the uncertainty of α value, which leads to variation of $E_F^{pre}$ values by several MeV or more, changes $\langle E \rangle$ by ~0.1 MeV [6, 9, 10]. The analysis of asymmetry of prompt fission neutron emission [5, 6, 9–16] with respect to the incident neutron momentum for $^{239}$Pu(*n, F*) [2, 17] and $^{235}$U(*n, F*) [8] reactions, might be applied for prediction of asymmetry of PFNS for $^{240}$Pu(*n, F*) reaction. Angle-averaged data on $^{240}$Pu(*n, F*) PFNS are compatible by the ENDF/B formatted data available at [12] before the paper [12] was published. **are α, α1 and $E_F^{pre}$, they**

Pre-fission neutrons influence the $^{239}$Pu(*n, F*) PFNS shape in the energy range of $E_n$~$E_{nnf}$–20 MeV as it was shown in [1, 6, 18]. They influence also the shape of TKE of fission fragments and products, prompt fission neutron multiplicity, fission fragment distributions and produce the step-like shape of observed fission cross section. The variation of observed average PFNS energies $\langle E \rangle$ in the vicinity of $^{240}$Pu(*n, xnf*) reaction thresholds is stronger than in case of $^{239}$Pu(*n, F*), due to larger contribution of exclusive spectra of $^{240}$Pu(*n, xnf*)$^{1,..x}$ neutrons [6]. Henceforth, the upper indices (1...*x*) notify the sequence of emitted *x* pre-fission neutrons. The amplitude of variations of $\langle E \rangle$ in case of $^{240}$Pu(*n, F*) [6, 9, 10] is consistent, up to $^{240}$Pu(*n*, 2*nf*) reaction threshold, with the preliminary angle-integrated data [12–15] for $\varepsilon$~0.8–10 Me  energy range.

Pre-fission neutrons in [2, 4, 8, 15–17] counted in coincidence with the fission fragments, without separation with respect to the fragment masses. The pre-fission relatively soft neutrons are emitted in a spherically symmetric way relative to the neutron beam momentum. The angular anisotropy of PFNS observed in $^{239}$Pu(*n, F*) [17], might be attributed to the non-compound emission of $^{239}$Pu(*n, nf*)$^1$ neutrons. The direction of the emission of (*n, nX*)$^1$ neutrons, as well as that of (*n, nγ*)$^1$, (*n, 2n*)$^1$, (*n, 3n*)$^1$ and (*n, nf*)$^1$, (*n, 2nf*)$^1$ and (*n, 3nf*)$^1$ neutrons, is correlated with the momentum of the incident neutrons. The direction of the neutrons emitted from the fission fragments correlates with the fission axis direction mostly. Both kinds of neutrons counted in coincidence with the fission fragments.

Anisotropy of NES of $^{239}$Pu+*n* interaction observed long ago [19]. The anisotropic contribution of double differential spectra of first neutron, relevant for the excitations of first residual nuclide of 0.5~6 MeV, is evidenced in double differential NES and mostly in the component of $^{239}$Pu(*n, nγ*)$^1$ reaction [5, 6, 9, 10]. However, the most investigated to define first neutron spectrum of (*n, nX*)$^1$ reaction are target nuclides $^{232}$Th or $^{238}$U, as shown in [5, 20]. Neutron emission spectra of $^{238}$U+*n* interaction are also strongly anisotropic. The experimental quasi-differential emissive neutron spectra for $^{235}$U+*n*, $^{238}$U+*n* and $^{239}$Pu+*n* interactions [21, 22] revealed once again the inadequacy of current NES modelling, envisaged in [20], and stimulated further efforts of NES modelling [23].

The level structures of $^{232}$Th, $^{238}$U and $^{240}$Pu are rather similar, they define the asymmetry of quasi-elastic peak of NES [5]. For example, direct excitation of $^{238}$U ground state band levels $J^{\pi}$ = 0$^+$, 2$^+$, 4$^+$, 6$^+$, 8$^+$ was accomplished in [24–26] within rigid rotator model, while that of $\beta$–bands of $K^{\pi}$ =0$^+$ and $\gamma$–bands of $K^{\pi}$ =2$^+$, octupole band of $K^{\pi}$ =0$^-$ at *U*=0~1.2 MeV was accomplished within soft deformable rotator. The net effect of these procedures is the adequate approximation of angular distributions of $^{232}$Th(*n, nX*)$^1$ and $^{238}$U(*n, nX*)$^1$ first neutron inelastic scattering in continuum which corresponds to *U*=1.2~6 MeV excitations for $E_n$ =1.2~20 MeV. The fictitious levels [27], as a substitute for collective levels inelastic scattering, we avoided. That approach is suitable to predict the NES of $^{240}$Pu+*n* interaction.

## II. PROMPT FISSION NEUTRON SPECTRA AND NEUTRON EMISSION SPECTRA

Prompt fission neutron spectra $S(\varepsilon, E_n, \theta)$ at angle $\theta$ relative to the incident neutron beam, is a superposition of exclusive spectra of (*n, xnf*)$^{1,\ldots x}$ pre-fission neutrons,–$\frac{d^2\sigma^k_{nxn}(\varepsilon, E_n, \theta)}{d\varepsilon d\theta}$ (*x*=0, 1, 2, 3; *k*=1,...,*x*), and spectra of prompt fission neutrons, emitted by fission fragments, $S_{A+1-x}(\varepsilon, E_n, \theta)$:

$$S(\varepsilon,E_n,\theta)=\tilde{S}_{A+1}(\varepsilon,E_n,\theta)+\tilde{S}_{A}(\varepsilon,E_n,\theta)+\tilde{S}_{A-1}(\varepsilon,E_n,\theta)+\tilde{S}_{A-2}(\varepsilon,E_n,\theta) \quad (1)$$

$$\tilde{S}_{A+1}(\varepsilon,E_n,\theta)=\nu_p^{-1}(E_n,\theta)\nu_{p1}(E_n)\cdot\beta_1(E_n,\theta)S_{A+1}(\varepsilon,E_n,\theta) \quad (2)$$

$$\tilde{S}_{A}(\varepsilon,E_n,\theta)=\nu_p^{-1}(E_n,\theta)\left[\nu_{p2}(E_n-\left\langle E_{nnf}(\theta)\right\rangle)\beta_2(E_n,\theta)S_A(\varepsilon,E_n,\theta)+\beta_2(E_n,\theta)\frac{d^2\sigma^1_{nnf}(\varepsilon,E_n,\theta)}{d\varepsilon d\theta}\right]$$

(3)

$$\tilde{S}_{A-1}(\varepsilon,E_n,\theta)=\nu_p^{-1}(E_n,\theta)\left\{\begin{array}{l}\nu_{p3}(E_n-B_n^A-\left\langle E^1_{n2nf}(\theta)\right\rangle-\left\langle E^2_{n2nf}(\theta)\right\rangle)\beta_3(E_n,\theta)S_{A-1}(\varepsilon,E_n,\theta)+\\ \beta_3(E_n,\theta)\left[\frac{d^2\sigma^1_{n2nf}(\varepsilon,E_n,\theta)}{d\varepsilon d\theta}+\frac{d^2\sigma^2_{n2nf}(\varepsilon,E_n,\theta)}{d\varepsilon d\theta}\right]\end{array}\right\} \quad (4)$$

$$\tilde{S}_{A-2}(\varepsilon,E_n,\theta)=\nu_p^{-1}(E_n,\theta)\left\{\begin{array}{l}\nu_{p4}(E_n-B_n^A-B_n^{A-1}-\left\langle E^1_{n3nf}(\theta)\right\rangle-\left\langle E^2_{n3nf}(\theta)\right\rangle-\left\langle E^3_{n3nf}(\theta)\right\rangle)\times\\ \beta_4(E_n,\theta)S_{A-2}(\varepsilon,E_n,\theta)+\\ \beta_4(E_n,\theta)\left[\frac{d^2\sigma^1_{n3nf}(\varepsilon,E_n,\theta)}{d\varepsilon d\theta}+\frac{d^2\sigma^2_{n3nf}(\varepsilon,E_n,\theta)}{d\varepsilon d\theta}+\frac{d^2\sigma^3_{n2nf}(\varepsilon,E_n,\theta)}{d\varepsilon d\theta}\right]\end{array}\right\} \quad (5)$$

In equations (1)-(5) $\tilde{S}_{A+1-x}(\varepsilon,E_n,\theta)$ is the contribution of ($x$+1)-th chance fission to the observed PFNS $S(\varepsilon,E_n,\theta)$, $\left\langle E^k_{nxnf}(\theta)\right\rangle$ – average energy of $k$–th neutron of ($n$, $xnf$) reaction with exclusive neutron spectrum, $\frac{d^2\sigma^k_{nxn}(\varepsilon,E_n,\theta)}{d\varepsilon d\theta}$, $k\leq x$. Spectra $S(\varepsilon,E_n,\theta)$, $S_{A+1-x}(\varepsilon,E_n,\theta)$ and $\frac{d^2\sigma^k_{nxn}(\varepsilon,E_n,\theta)}{d\varepsilon d\theta}$ are normalized to unity. Index ($x$+1) denotes the fission chance of $^{241-x}$Pu nuclides after emission of $x$ pre-fission neutrons, $\beta_x(E_n,\theta)=\sigma_{n,xnf}(E_n,\theta)/\sigma_{n,F}(E_n,\theta)$ – relative contribution of ($x$+1)–th fission chance to the observed fission cross section $\sigma_{n,F}(E_n,\theta)$, $\nu_p(E_n,\theta)$ is the average number of prompt fission neutrons, $\nu_{px}(E_{nx})$ – average number of prompt fission neutrons, emitted by the fragments of fission of $^{241-x}$Pu nuclides. Spectra of prompt fission neutrons, emitted from the fragments, $S_{A+1-x}(\varepsilon,E_n,\theta)$, as proposed in [28], were approximated by the sum of two Watt distributions [29] with different temperatures, the temperature of the light fragment being higher.

Modelling the angular distribution for the exclusive spectra of pre-fission neutrons $^{239}$Pu$(n,xnf)^{1,\dots x}$ [18], we reproduced measured PFNS ratios $R(\varepsilon,E_n,\Delta\theta,\Delta\theta^1)=\left\langle S(\varepsilon,E_n,\Delta\theta)\right\rangle_{\Delta E_n}/\left\langle S(\varepsilon,E_n,\Delta\theta^1)\right\rangle_{\Delta E_n}$ in angular ranges $\Delta\theta$ ~35°–40° and $\Delta\theta^1\approx130^o-140^o$ for wide energy range of $\Delta E_n$ ~15–17.5 MeV [17]. The ratios we extracted from the measured data, shown on Fig.3 of paper [17]. The ratios $R(\varepsilon,\Delta E_n,\Delta\theta,\Delta\theta^1)$ demonstrated on Fig. 1. PFNS representation alternative to that shown on Fig.3 in paper [17], as a ratio $R^{\exp}=S(\varepsilon,E_n\approx15-17.5,\Delta\theta)/S(\varepsilon,E_n\approx15-17.5,\Delta\theta^1)$ for $\Delta\theta$ ~35°–40° (forward direction) and $\Delta\theta^1$ ~130°–140° (backward direction) is virtually independent upon the normalizations adopted in [17]. In [5, 11, 30, 31] it was shown, that the calculated ratios $R(\varepsilon, E_n, \Delta\theta, \Delta\theta^1)$ for $^{232}$Th(*n, F*) and $^{238}$U(*n, F*) strongly depend on the $^{232}$Th(*n, xnf*) $^{1,\ldots x}$ and $^{238}$U(*n, xnf*) $^{1,\ldots x}$ reactions relative contributions to the respective observed PFNS and fission cross sections. Anomalous values of $R(\varepsilon, E_n, \Delta\theta, \Delta\theta^1)$ were predicted for the $^{232}$Th(*n, F*) reaction PFNS [20]. Blind application of the approach, described in [1, 5, 11, 18, 20, 30, 31], to predict angle-integrated PFNS of $^{240}$Pu(*n, F*) and then the asymmetry of pre-fission neutrons, produces $R(\varepsilon, E_n, \Delta\theta, \Delta\theta^1)$, shown on Fig. 1.

The measured ratios of average energies of PFNS $\langle E(\theta \approx 37.5^o)\rangle / \langle E(\theta^1 \approx 135^o)\rangle$ [17], i.e. of average energies of prompt fission neutrons $\langle E \rangle$ for neutrons counted at angular intervals $\Delta\theta$ ~35°–40° and $\Delta\theta^1$~130°–150° in $E_n$~1–20 MeV incident neutron energy range were reproduced for $^{239}$Pu(*n, F*) and predicted for $^{240}$Pu(*n, F*) also (Fig. 2). The main factor for the observed features of PFNS, like ratios $R(\varepsilon, E_n, \Delta\theta, \Delta\theta^1)$ and $\langle E(\theta \approx 37.5^o)\rangle / \langle E(\theta^1 \approx 135^o)\rangle$, is the excitation energy of fissioning Pu nuclides emerging after *x* pre-fission neutron emission. Average energies for the angle-integrated PFNS of $^{240}$Pu(*n, F*) for $\varepsilon$~0.8–10 MeV [12–14] are compared with calculations, described in [6, 9, 10], on Fig. 3. Correlation of the, average energy of 1st neutron of (*n, nf*) reaction $\langle E^1_{nnf}(\theta)\rangle$ with $\langle E \rangle$ of PFNS is evident. At higher incident neutron energies the influence of $\langle E^{1,2}_{n2nf}(\theta)\rangle$ – average energies of 1st and 2nd neutrons of $^{240}$Pu(*n,2nf*) reaction is weaker, than observed in [12–14].

Since the energy and angular intervals in measurements of [17] are rather wide, the number of detected prompt fission neutrons are calculated in a similar fashion as NES. Double differential NES defined as

$$
\begin{aligned}
\frac{d^2\sigma(\varepsilon,E_n,\theta)}{d\varepsilon d\theta} = \frac{1}{2\pi}\Big[ & \nu_p(E_n,\theta)\sigma_{nF}(E_n,\theta)S(\varepsilon,E_n,\theta) + \sigma_{nn\gamma}(\varepsilon,E_n,\theta)\frac{d^2\sigma^1_{nn\gamma}(\varepsilon,E_n,\theta)}{d\varepsilon d\theta} + \\
& \sigma_{n2n}(\varepsilon,E_n,\theta)\left(\frac{d^2\sigma^1_{n2n}(\varepsilon,E_n,\theta)}{d\varepsilon d\theta} + \frac{d^2\sigma^2_{n2n}(\varepsilon,E_n,\theta)}{d\varepsilon d\theta}\right) + \\
& \sigma_{n3n}(\varepsilon,E_n,\theta)\left(\frac{d^2\sigma^1_{n3n}(\varepsilon,E_n,\theta)}{d\varepsilon d\theta} + \frac{d^2\sigma^2_{n3n}(\varepsilon,E_n,\theta)}{d\varepsilon d\theta} + \frac{d^2\sigma^3_{n3n}(\varepsilon,E_n,\theta)}{d\varepsilon d\theta}\right) + \\
& \sum_q \frac{d\sigma_{nn\gamma}(\varepsilon,E_q,E_n,\theta)}{d\theta} G(\varepsilon,E_q,E_n,\Delta_\theta)\Big],
\end{aligned}
\qquad (6)
$$

$$
G(\varepsilon,E_q,E_n,\Delta_\theta) = \frac{2}{\Delta_\theta\sqrt{\pi}}\exp\left\{-\left[\frac{\varepsilon-(E_n-E_q)}{\Delta_\theta}\right]^2\right\}. \qquad (7)
$$

NES in Eq. (6) is a superposition of prompt fission neutron spectra $S(\varepsilon, E_n, \theta)$, exclusive neutron spectra of (*n, n* $\gamma$ )$^1$, (*n*, 2*n*)$^{1,2}$ and (*n*, 3 *n*)$^{1,2,3}$ reactions, $\frac{d^2\sigma^k_{nxn}(\varepsilon,E_n,\theta)}{d\varepsilon d\theta}$, normalized to unity, and spectra of elastic and inelastic scattered neutrons, followed by excitation of collective levels $E_q$ of $^{240}$Pu nuclide, $\frac{d^2\sigma_{nn\gamma}(\varepsilon,E_q,E_n,\theta)}{d\varepsilon d\theta}$. $G(\varepsilon, E_q, E_n, \Delta_\theta)$–resolution ($\Delta_\theta$) function, which depends on $E_n$ and may weakly depend on $\theta$ and $E_q$. The calculated NES are normalized using average prompt fission neutron multiplicity, (*n, F*) and (*n, xn*) cross sections.

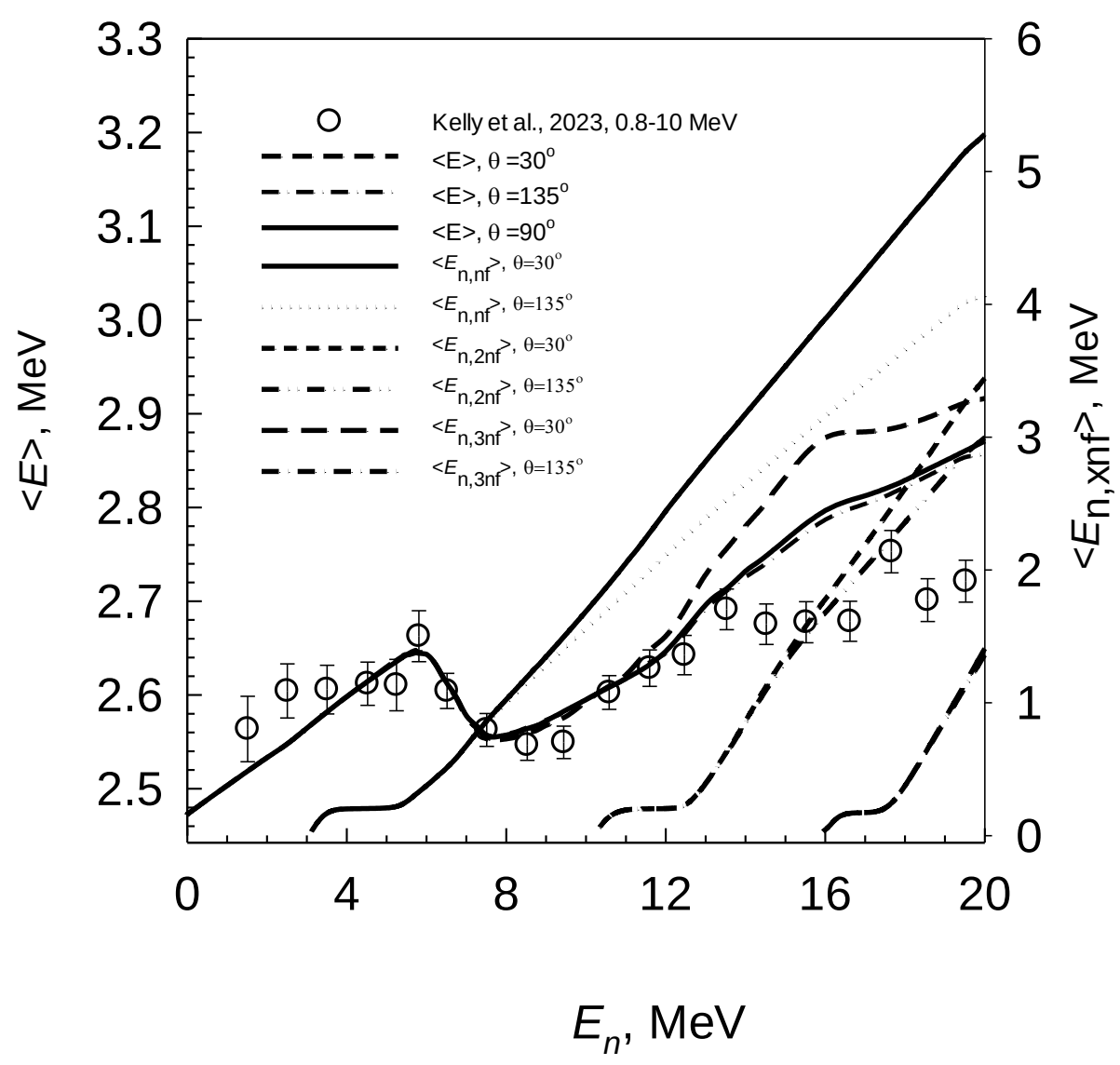

Fig. 3 PFNS $\langle E\rangle$ for $^{240}$Pu(*n, F*), $\varepsilon$~0.8–10 MeV: ○–[12]; full line, –$\langle E(90^o)\rangle$; dashed line – $\langle E(30^o)\rangle$; dash-dotted line–$\langle E(135^o)\rangle$; solid line – $\langle E_{n,nf}(\theta\approx 30^o)\rangle$; dotted line–$\langle E_{n,nf}(\theta\approx 135^o)\rangle$; dashed line–$\langle E_{n,2nf}(\theta\approx 30^o)\rangle$; dashed – double dotted line–$\langle E_{n,2nf}(\theta\approx 135^o)\rangle$; long dashed line–$\langle E_{n,3nf}(\theta\approx 30^o)\rangle$; dash dotted line–$\langle E_{n,3nf}(\theta\approx 135^o)\rangle$.

The part of inclusive anisotropic double differential spectrum of first neutron relevant for the excitations close to the fission barrier value of $^{240}$Pu nuclide (corresponds to second chance fission reaction), will be pronounced in exclusive spectra of $(n, nf)^1$, $(n, 2nf)^1$ and $(n, 2n)^1$ at $E_n \gtrsim 12$ MeV, at various pre-fission neutron emission angles $\theta$. Angular distribution of pre-fission neutrons in [17] was extracted from the observed PFNS of $^{239}$Pu(*n, F*) by subtracting the post-fission neutron spectrum, which was estimated in rather approximate manner, due to adopted normalizations of PFNS to the equal number of fission events for any angle $\theta$.

## III. FISSION CROSS SECTION $\sigma_{n,F}(E_n)$ AND PROMPT FISSION NEUTRON MULTIPLICITY $\nu_p(E_n)$

Contributions $\beta_x(E_n,\theta)$ of $x$–th fission chance (*n, xnf*) to the observed fission cross section $\sigma_{n,F}(E_n)$ in Eqs. (2) – (5) is about the main factor influencing the shape of PFNS. Figure 4 shows the values $\beta_1(E_n)$ and $\beta_2(E_n)$ of 1st and 2nd fission chances to the observed fission cross sections $\sigma_{n,F}(E_n)$ of $^{239}$Pu(*n, F*) and $^{240}$Pu(*n, F*) reactions. Contributions $\beta_1 = \sigma_{n,f}/\sigma_{n,F}$ and $\beta_2 = \sigma_{n,nf}/\sigma_{n,F}$ of [32] were estimated by the analysis of prompt fission neutron multiplicity distributions. They are quite different from $\beta_x(E_n)$ values, used in [1, 6, 18, 30, 31, 33–36]. The $\beta_x(E_n)$ estimates of [32] seem to be rather unstable with respect to the experimental uncertainties. The experimental points $\beta_x(E_n)$ (Fig. 4) were obtained renormalizing the data of [32] as $\tilde{\beta}_2(E_n) = 0.7\beta_2(E_n)$. When renormalized, $\tilde{\beta}_1(E_n)$ and $\tilde{\beta}_2(E_n)$ appear quite consistent with those used in [1, 6, 18, 30, 31, 33–36], especially near reaction thresholds of $^{239}$Pu(*n, xnf*) reaction. Our estimates of $\beta_1(E_n)$ and $\beta_2(E_n)$ of $^{239}$Pu(*n, F*) are much more grounded, since they allow to reproduce available observed PFNS at $E_{nnf} \lesssim E_n \lesssim 20$

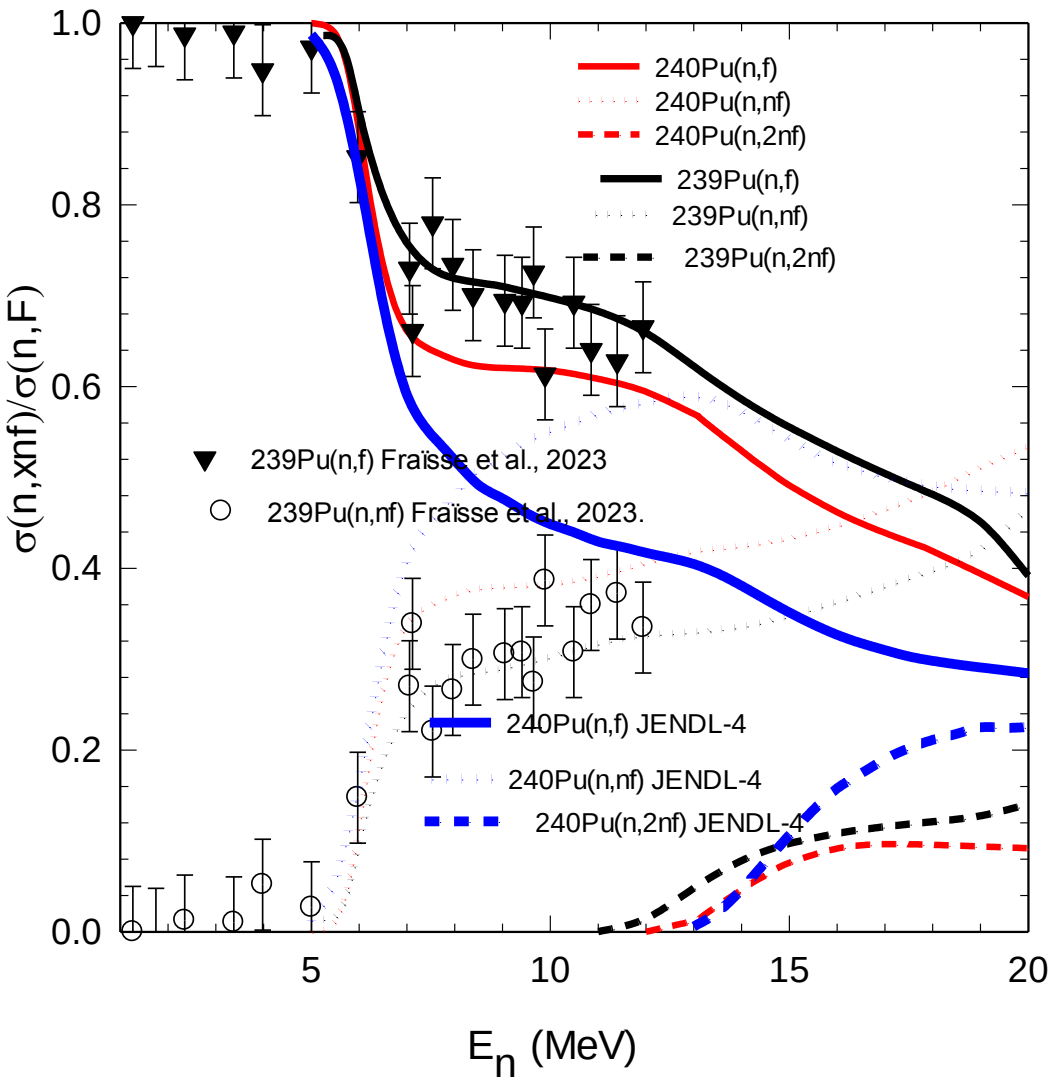

Fig.4. Ratios of partial components (*n,xnf*) to the observed fission cross section $^{239,240}$Pu(*n, F*) and $^{239,240}$Pu(*n, F*). Red curves — $^{240}$Pu (*n,xnf*); black curves — $^{239}$Pu(*n,xnf*); blue curves — $^{240}$Pu (*n,xnf*) of [32]; ○ — [32]; ▼ — [32].

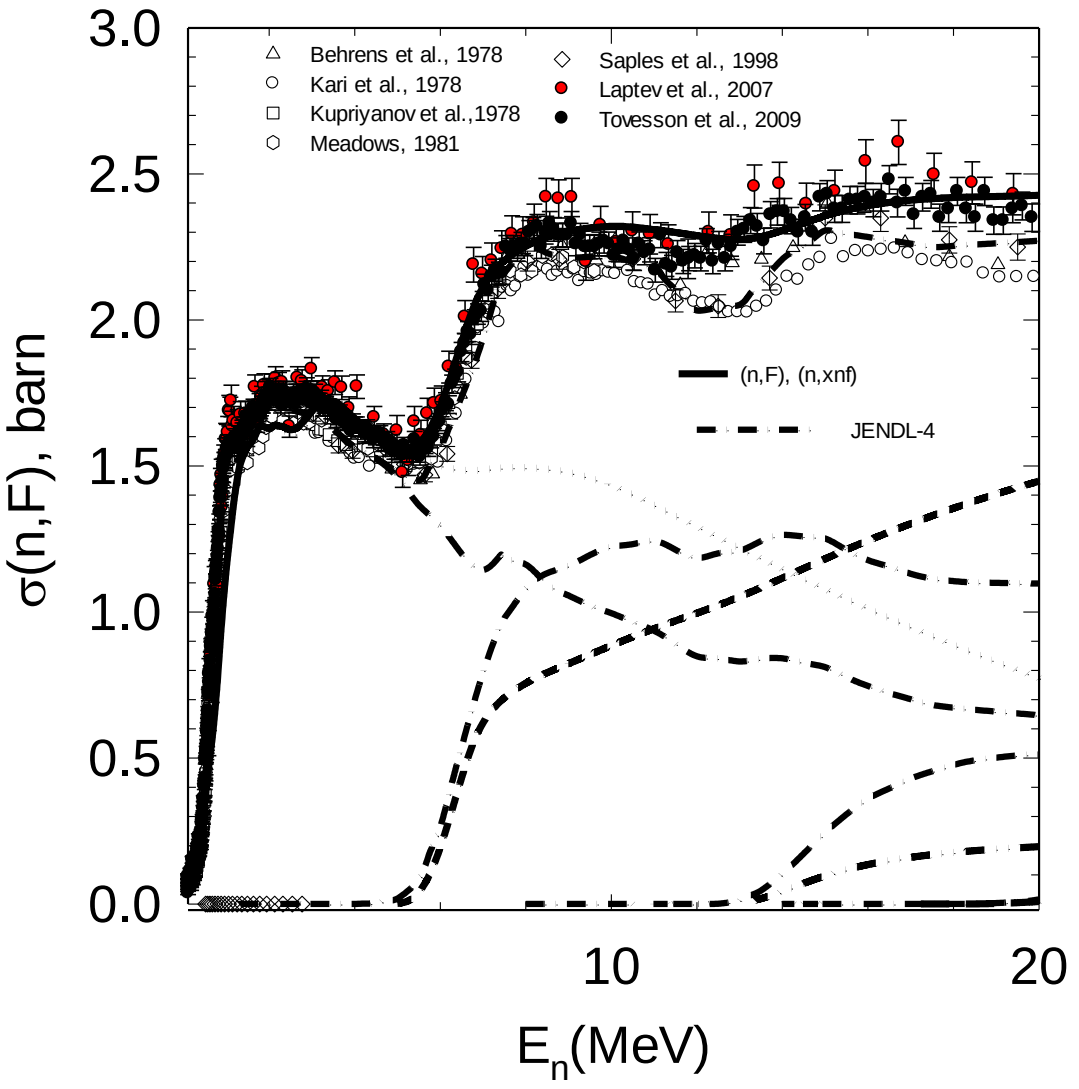

Fig.5. Partial $^{240}$Pu (*n,xnf*) components of $^{240}$Pu(*n, F*) ; ○ — [37]; ◊ — [38]; ◦ — [39]; □ — [40]; Δ — [41]; ● — [42]; ● — [43]; solid line — $^{240}$Pu(*n, F*); dotted line — $^{240}$Pu(*n, f*); dashed line — $^{240}$Pu(*n, nf*); dashed double dotted line — $^{240}$Pu(*n, 2nf*); ]; dash dotted curves — $^{240}$Pu (*n,xnf*) of [44];

MeV using $\tilde{S}_{240}(\varepsilon, E_n)$, $\tilde{S}_{239}(\varepsilon, E_n)$ and $\tilde{S}_{238}(\varepsilon, E_n)$ contributions of $^{239}$Pu(*n,xnf*) reactions. Calculated values of $\beta_2(E_n)$ for $^{240}$Pu(*n, F*) are systematically higher than $\beta_2(E_n)$ of $^{239}$Pu(*n, F*), the $\beta_2(E_n)$ for $^{240}$Pu(*n, F*) are incompatible with relevant estimates of [32] (se Fig. 4).

The observed fission cross section of $^{240}$Pu(*n, F*) was calculated in a Hauser-Feshbach formalism as

$$\sigma_{n,F}(E_n) = \sigma_{n,f}(E_n) + \sum_{x=1}^{3} \sigma_{n,xnf}(E_n) \quad , \qquad (8)$$

and compared with measured data [37–43] on Fig. 5. The (*n, xnf*) reaction contributions are defined by the fission probability $P_{f(241-x)}^{J\pi}(E)$ of $^{241-x}$Pu nuclides:

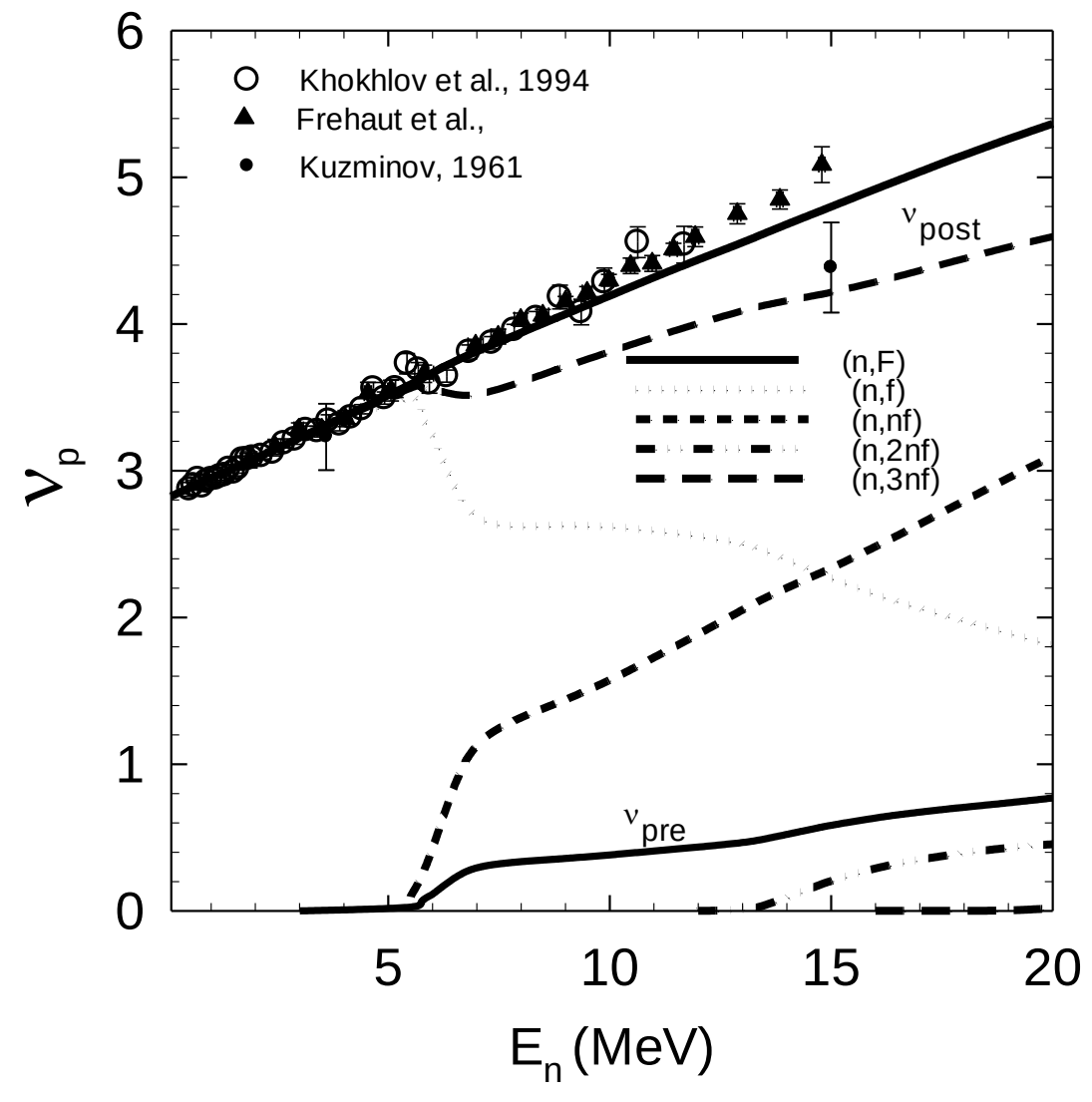

Fig. 6. Average number of prompt fission neutrons of $^{240}$Pu(*n,F*) and its partial components: solid line — $^{240}$Pu; dotted line — $^{240}$Pu(*n, f*); dashed line — $^{240}$Pu(*n, nf*); dashed double dotted line — $^{240}$Pu(*n*, 2*nf*); dashed line — $\nu_{post}$ ; solid line — $\nu_{pre}$ ; ○ — [45]; ▲ — [46]; • — [47].

$$\sigma_{n,xnf}\left(E_n\right)=\sum_{J\pi}^{J}\int_{0}^{U_x}W^{J\pi}_{A+1-x}(U)P^{J\pi}_{f(A+1-x)}(U)dU\,, \qquad (9)$$

here $W^{J\pi}_{A+1-x}(U)$ is the population of excited states of $^{241-x}$Pu nuclides with excitation energy *U,* after emission of *x* pre-fission neutrons. Calculated cross sections of $^{240}$Pu(*n, xnf*) reactions are much different from estimates of [44] (see Fig. 5). If the latter estimates of $^{240}$Pu(*n,xnf*) contributions would be used for calculation of PFNS, large discrepancies with present calculations would be due to differing contribution of $^{240}$Pu(*n, nf*)$^1$ and $^{240}$Pu(*n,* 2*nf*)$^{1,2}$ pre-fission neutrons and neutrons, emitted from the fragments $S_{241-x}(\varepsilon,E_n,\theta)$.

Observed average number of prompt fission neutrons $\nu_p(E_n)$ defined as superposition of pre-fission neutrons and neutrons, coming from the fission fragments of $^{241-x}$Pu nuclides, cooled down by the pre-fission neutron emission is

$$\nu_{\rm p}(E_n)=\nu_{post}+\nu_{pre}=\sum_{x=0}^{3}\nu_{px}(E_{nx})+\sum_{x=1}^{3}x\cdot\beta_{x+1}(E_n)\,. \qquad (10)$$

The post-fission, $\nu_{post}(E_n)$ and pre-fission $\nu_{pre}(E_n)$, components as well as (*n, xnf*) partial contributions of $\nu_p(E_n)$ were obtained by consistent description of $\nu_p(E_n)$ and observed fission cross sections at $E_n$ <20 MeV (see Fig. 6).

Average prompt fission neutron number $\nu_p(E_n)$ of $^{240}$Pu(*n, F*) was measured in [45–47]. Partial average neutron multiplicities $\nu_{px}(E_{nx})$ define relative contributions of pre-fission neutrons with spectra $d\sigma^k_{nxnf}/d\varepsilon$ and prompt neutrons, emitted from fission fragments, with spectra $S_{A+1-x}(\varepsilon,E_n)$, see Eqs. (1)-(5). To calculate $\nu_{\rm p}(E_n)$ values when $E_n>E_{nnf}$, the data on $\nu_{\rm p}(E_n)$ for $^{241-x}$Pu(*n, F*) nuclides at $E_n<E_{nnf}$ should be used, when available. The partial contributions $\tilde{S}_{A+1-x}(\varepsilon, E_n)$ to the PFNS correlate with partial contributions $\nu_{px}(E_{nx})$ to the $\nu_p(E_n)$. Figure 6 compares the model calculation of $\nu_p(E_n)$ (Eq. (10)) with measured data [45–47]. Partial contributions of $^{240}$Pu(*n, F*), i.e., $^{240}$Pu(*n, f*) $^{240}$Pu(*n, nf*) and $^{240}$Pu(*n, 2nf*) reactions are shown, as well as lumped contributions of six pre-fission neutrons of $^{240}$Pu$(n, nf)^1$, $^{240}$Pu$(n, 2nf)^{1,2}$ and $^{240}$Pu$(n, 3nf)^{1,2,3}$. Contributions of pre-fission neutrons, $\nu_{pre}(E_n)$, and post-fission neutrons, emitted from relevant fission fragments, $\nu_{post}(E_n)$ are also shown. Partial contribution of $^{240}$Pu(*n, nf*) reaction influences but still only weakly smooth energy dependence of $\nu_p(E_n)$ around $^{240}$Pu(*n, nf*) reaction threshold $E_{nnf}$. The strong bump may appear in $\sigma_{n,F}(E_n)$ and $\nu_p(E_n)$ of $^{232}$Th(*n, F*) or $^{230}$Th(*n, F*), when $E_n > E_{nnf}$ [48, 49]. That happens because of net influence of wide (~1.5 MeV) pairing gap at saddle deformations, relatively high fission barrier of $^{232}$Th or $^{230}$Th nuclides and rather high thresholds of $^{232}$Th(*n*, 2*n*) or $^{230}$Th(*n*, 2*n*) reactions. In case of $^{238}$U(*n, F*) or $^{240}$Pu(*n,F*) only small steps are observed in $\sigma_{n,F}(E_n)$ [50]. The values of $\nu_{post}(E_n)$ and $\nu_{pre}(E_n)$ of $^{240}$Pu(*n, F*) and $^{239}$Pu(*n, F*) are influenced by the values of $\beta_x(E_n) = \sigma_{n,xnf} / \sigma_{n,F}$ mostly (see Fig. 6).

## IV. EXCLUSIVE PRE-FISSION NEUTRON SPECTRA $d^2\sigma^k_{nxnf}/d\varepsilon d\theta$

To calculate contributions of pre-fission neutrons with spectra $d^2\sigma^k_{nxnf}/d\varepsilon d\theta$ to the PFNS $S(\varepsilon, E_n, \theta)$ (Eq. 1) the inclusive neutron emission spectrum of $(n,nX)^1$ reaction, $\frac{d^2\sigma^1_{nnX}(\varepsilon, E_n, \theta)}{d\varepsilon d\theta}$, should be defined. It is the sum of compound and weakly dependent on emission angle pre-equilibrium components, that procedure followed for years. To reproduce angular asymmetry of NES a phenomenological function proposed in [36], modelling energy and angle dependence of NES due to the first neutron inelastic scattering in continuum:

$$\frac{d^2\sigma^1_{nnX}(\varepsilon, E_n, \theta)}{d\varepsilon d\theta} \approx \frac{d^2\tilde{\sigma}^1_{nnX}(\varepsilon, E_n, \theta)}{d\varepsilon d\theta} + \sqrt{\frac{\varepsilon}{E_n}} \frac{\omega(\theta)}{E_n - \varepsilon} \qquad (11)$$

$$\omega(\theta) = 0.4\cos^3(\theta) + 0.16 \qquad (12)$$

Angle-averaged function $\omega(\theta)$, i.e., $\langle\omega(\theta)\rangle_\theta$ for angles $\theta_2 - \theta_1 = 135^o - 30^o$ [17], is approximated as $\langle\omega(\theta)\rangle_\theta \approx \omega(90^o)$, then angle-integrated spectrum is

$$\frac{d\sigma^1_{nnX}(\varepsilon, E_n)}{d\varepsilon} \approx \frac{d\tilde{\sigma}^1_{nnX}(\varepsilon, E_n)}{d\varepsilon} + \sqrt{\frac{\varepsilon}{E_n}} \frac{\langle\omega(\theta)\rangle_\theta}{E_n - \varepsilon}. \qquad (13)$$

To retain the flux conservation in cross section and spectra calculations the compound reaction cross section renormalized to account for extra semi-direct neutron emission:

$$\sigma_c(E_n) = \sigma_a(E_n)(1 - q - \tilde{q}). \qquad (14)$$

Inclusive emission spectrum of the interaction $(n,nX)^1$, $q$–ratio of pre-equilibrium neutrons in a standard pre-equilibrium model [51],

$$\frac{d\tilde{\sigma}^1_{nnX}(\varepsilon,E_n)}{d\varepsilon}=\sum_{J,\pi}W^{J\pi}_{240}(E_n-\varepsilon,\theta)\,, \qquad (15)$$

depends on the fission probability of $^{241}$Pu nuclide, $W^{J\pi}_{240}(E_n-\varepsilon,\theta)$ is the population of residual nuclide $^{240}$Pu states with spin/parity $J^{\pi}$ and excitation energy $U=E_n-\varepsilon$, after first neutron emission at angle $\theta$. It is used to define the exclusive spectra of each partial reaction, $d\sigma^k_{nxnf}/d\varepsilon$ or $d\sigma^k_{nxn}/d\varepsilon$, as well as $d^2\sigma^k_{nxnf}/d\varepsilon d\theta$ or $d^2\sigma^k_{nxn}/d\varepsilon d\theta$ in a STAPRE code framework [51]. Henceforth, the indexes $J^{\pi}$ in fission, $\Gamma_f$, neutron $\Gamma_n$ and total $\Gamma$ widths described in [48-50], as well as relevant summations, omitted. The angular dependence of partial width, calculated with spin and parity conservation, is due to dependence of excitation energy of residual nuclides on the emission angle $\theta$ of first neutron. The exclusive spectra of pre-fission $(n, nf)^1$ neutrons are

$$\frac{d^2\sigma^1_{nnf}(\varepsilon,E_n,\theta)}{d\varepsilon d\theta}=\frac{d^2\sigma^1_{nnX}(\varepsilon,E_n,\theta)}{d\varepsilon d\theta}\frac{\Gamma^{240}_f(E_n-\varepsilon,\theta)}{\Gamma^{240}(E_n-\varepsilon,\theta)}. \qquad (16)$$

Exclusive first neutron spectra for the reaction $(n, 2nf)$, $(n, 2nf)^1$, is defined as:

$$\frac{d^2\sigma^1_{n2nf}\left(\varepsilon,E_n,\theta\right)}{d\varepsilon d\theta}=\int_0^{E-B_n^{240}}\frac{d^2\sigma^1_{n2nX}\left(\varepsilon,E_n,\theta\right)}{d\varepsilon d\theta}\frac{\Gamma^{239}_f(E_n-B_n^{240}-\varepsilon-\varepsilon_1)}{\Gamma^{239}(E_n-B_n^{240}-\varepsilon-\varepsilon_1)}d\varepsilon_1\,. \qquad (17)$$

Inclusive first neutron spectra of $(n, 2nX)$ reaction, $(n,2nX)^1$, is defined by the neutron spectrum of $(n,nX)^1$ and neutron emission probability of nuclide $^{240}$Pu as:

$$\frac{d^2\sigma^1_{n2nX}\left(\varepsilon,E_n,\theta\right)}{d\varepsilon d\theta}=\frac{d^2\sigma^1_{nnX}\left(\varepsilon,E_n,\theta\right)}{d\varepsilon d\theta}\frac{\Gamma^{240}_n(E_n-\varepsilon,\theta)}{\Gamma^{240}(E_n-\varepsilon,\theta)}\,. \qquad (18)$$

Phenomenological approach of Eqs. (1)–(18) enables to reproduce angular dependent NES in case of $^{232}$Th+$n$, $^{235}$U+$n$, $^{238}$U+$n$ and $^{239}$Pu+$n$ interactions [5].

Angle-integrated exclusive spectra of pre-fission $^{240}$Pu$(n, nf)^1$ neutrons are represented as

$$\frac{d\sigma^1_{nnf}(\varepsilon,E_n)}{d\varepsilon}=\frac{d\sigma^1_{nnX}(\varepsilon,E_n)}{d\varepsilon}\frac{\Gamma^{240}_f(E_n-\varepsilon)}{\Gamma^{240}(E_n-\varepsilon)}. \qquad (19)$$

Figure 7 shows the PFNS of $^{240}$Pu$(n,nf)$ in the incident neutron energy range $E_n$ ~5.75–7.5 MeV with a step of $E_n$ ~0.25 MeV. Interplay of pre-fission neutrons and prompt fission neutrons is quite noticeable. Preliminary PFNS data of [14, 15] in the energy range $E_n$ ~7–8 MeV and calculated PFNS at $E_n$ ~7.5 MeV [9, 10] are consistent with each other. However, it seems the cut-off energy $E_{nnf1}$ of pre-fission $^{240}$Pu$(n, nf)^1$ neutrons [14, 15] is slightly shifted to the higher energies $\varepsilon$ in comparison with calculations of [9, 10]. PFNS, calculated at $E_n$ ~5.75–7.5 MeV clearly demonstrate how the pre-fission neutrons influence the PFNS shapes. In fact, $\langle E\rangle$ attains the value observed at $E_n$ ~5 MeV only at $E_n$ ~12 MeV (see Fig. 3). The data of [12] at $E_n$ ~5.5–6.0 MeV and $E_n$ ~7.0–8.0 MeV reveal the influence of pre-fission $^{240}$Pu($n$, $nf$)[1] neutrons on the PFNS shape at $\varepsilon \gtrsim E_{nnf1}$. Due to cooling down of nuclides $^{240}$Pu after pre-fission neutron emission in the reaction $^{240}$Pu($n$, $nf$), the PFNS at $E_n$ ~5.75 MeV increases more steeply than relevant component of PFNS at $E_n$ ~7.5 MeV. The comparison of calculated $^{240}$Pu($n$, $nf$) reaction contribution with the quasi-experimental data [12], obtained by subtracting the calculated $^{240}$Pu($n$, $f$) contribution at $E_n$ ~7.5 MeV from the relevant PFNS data of [12] shows good consistency of shapes both at $\varepsilon \lesssim E_{nnf1}$ and $\varepsilon \gtrsim E_{nnf1}$. At $E_n$ ~5.75 MeV the pre-fission neutrons are not observed, but they influence the PFNS contribution $S_A(\varepsilon, E_n, \theta)$ of the second chance fission due to the cooling down of nuclides $^{240}$Pu in the reaction $^{240}$Pu($n,nf$). The comparison of calculated $^{240}$Pu($n,nf$) reaction contribution with the quasi-experimental data [12], obtained by subtracting the calculated $^{240}$Pu($n,f$) contribution at $E_n$ ~5.75 MeV from PFNS data of [12] shows perfect consistency of shapes at $\varepsilon \gtrsim E_{nnf1}$.

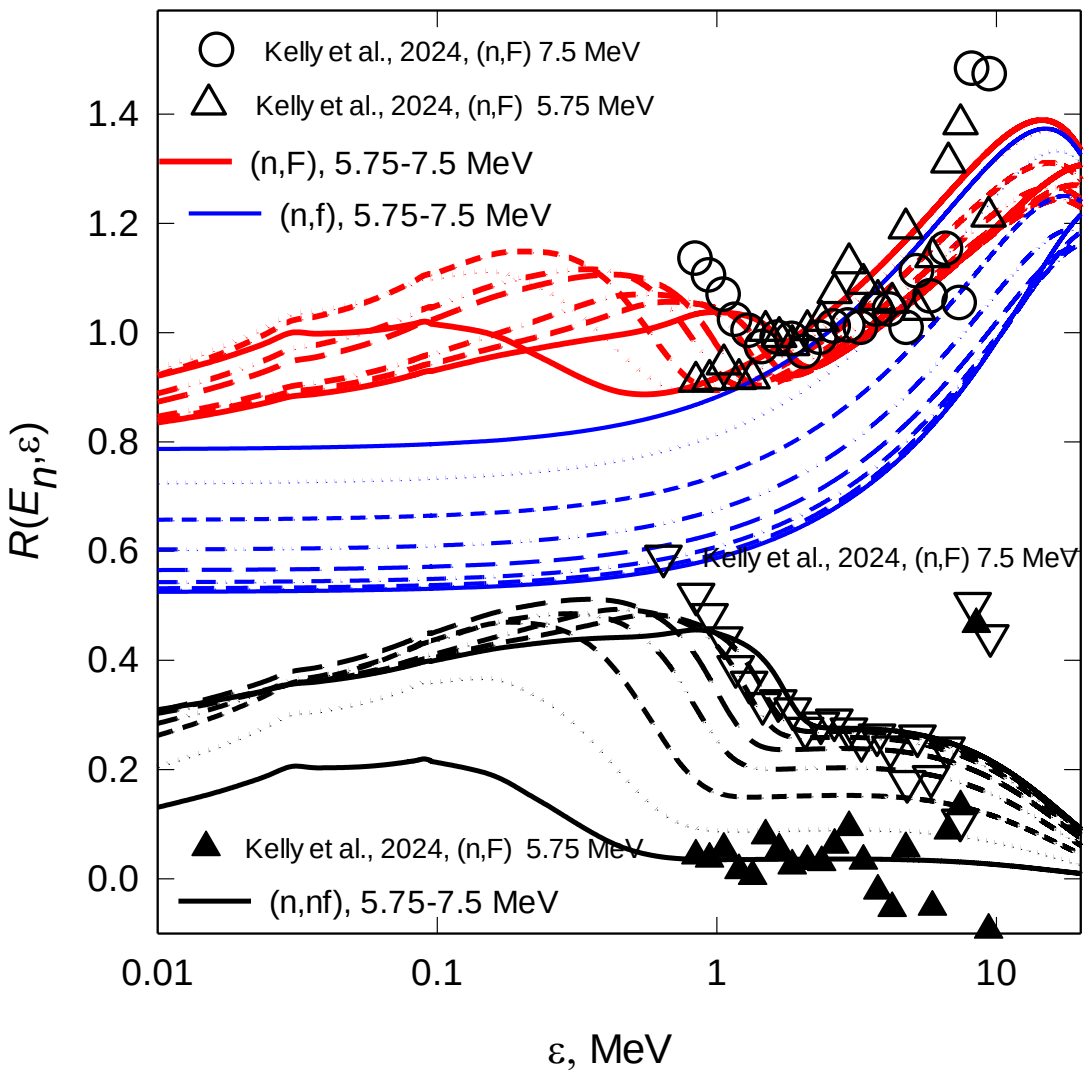

Fig. 7. PFNS of $^{240}$Pu($n,F$) at $E_n$ ~ 5.75 —7.5 MeV with step of 0.25 MeV and its partial components: red lines — $^{240}$Pu($n$, $F$); blue lines — $^{240}$Pu($n$, $f$); black lines — $^{240}$Pu($n$, $nf$); ○ — $^{240}$Pu($n$, $F$) [12], 7.5 MeV; ∇ — $^{240}$Pu($n$, $nf$) [12], 7.5 MeV; Δ — $^{240}$Pu($n$, $F$) [12], 5.75 MeV; ▲ — $^{240}$Pu($n$, $nf$) [12], 5.75 MeV.

## V. ANGULAR DISTRIBUTION OF $^{240}$Pu($n,xnF$) PRE-FISSION NEUTRONS

Exclusive pre-fission neutron spectra of $^{240}$Pu($n$, $nf$)[1] comprise small part of inclusive ($n,nX$)[1] spectrum, nonetheless it might be argued that they are responsible for the angular dependence of PFNS with respect to the incident neutron beam [10]. Figure 8 shows the partial contributions of $^{240}$Pu($n$, $f$), $^{240}$Pu($n$, $nf$) and $^{240}$Pu($n$, 2$nf$) contributions to the $^{240}$Pu($n$, $F$) PFNS at $E_n$ ~13.5 MeV at $\theta$ ~90° and $\theta$ ~30°. Even at this low incident energy PFNS shape for forward emission is much harder than that corresponding to pre-fission neutron emission at $\theta$ ~90°. The hardening of $\langle E \rangle$ at $\theta$ ~30° is due to increase of the second chance fission contribution and hardening of pre-fission neutron spectra $\dfrac{d^2\sigma^1_{nnf}(\varepsilon, E_n, \theta)}{d\varepsilon d\theta}$. Figure 9 compares the partial contributions of $^{240}$Pu($n$, $f$), $^{240}$Pu($n$, $nf$) and $^{240}$Pu($n$, 2$nf$) contributions to $^{240}$Pu($n$, $F$) PFNS at $E_n$ ~13.5 MeV at $\theta$ ~90° and $\theta$ ~30° with PFNS of $^{238}$U($n$, $F$). Even at this low incident energy PFNS shape for forward emission is much harder than that corresponding to pre-fission neutron emission at $\theta$ ~30°. Contribution of $^{238}$U($n$, 2$nf$) reaction to the PFNS of $^{238}$U($n$, $F$) is much higher than in case of $^{240}$Pu($n$, $F$) PFNS. Even at this low incident energy PFNS shape for forward emission is much harder than that corresponding to pre-fission neutron emission at $\theta$ ~30°.

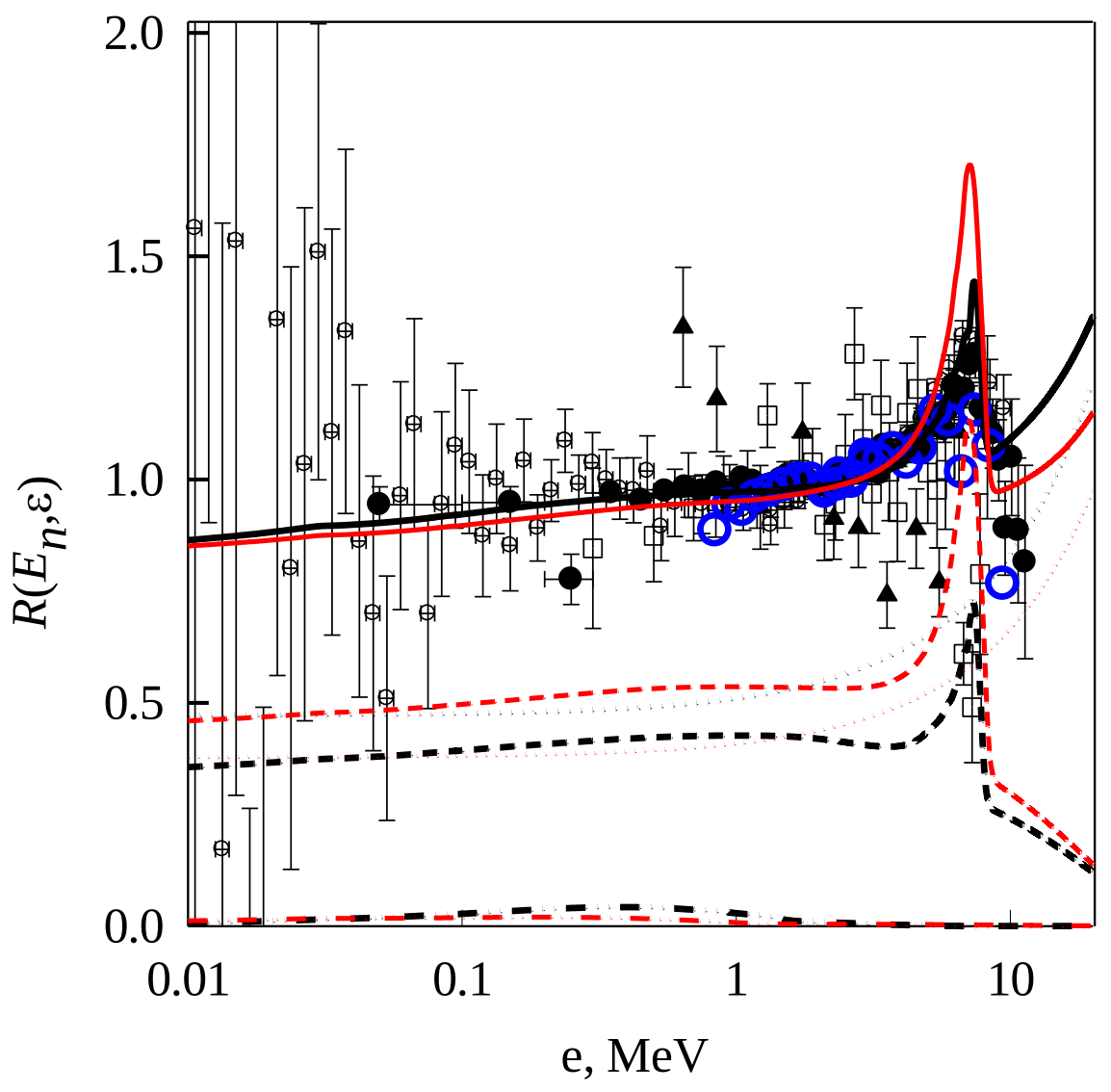

Fig.8. Ratios of partial components of PFNS of $^{240}$Pu($n$, $F$) and $^{239}$Pu($n$, $F$) at $E_n$ = 13.5 MeV relative to Maxwellian type distribution with $T$ = 1.4241 MeV: —— – $^{240}$Pu($n$, $F$); • • • — $^{240}$Pu($n$, $F$); – – – – $^{240}$Pu($n$, $nf$); – –• • – – – $^{240}$Pu($n$, 2$nf$); —— – $^{240}$Pu ($n$, $F$), $\theta$=30°; • • • – $^{240}$Pu($n$, $F$), $\theta$=30°; – – – – $^{240}$Pu($n$, $nf$) ), $\theta$=30°; – –• • – – – $^{240}$Pu($n$, 2$nf$) ), $\theta$=30°; ○ — $^{239}$Pu($n$, $F$) [2]; ● — $^{239}$Pu($n$, $F$) [3]; □— $^{239}$Pu($n$, $F$) [52]; ▲— $^{239}$Pu($n$, $F$) [55]; ○ — $^{240}$Pu($n$, $F$) [1].

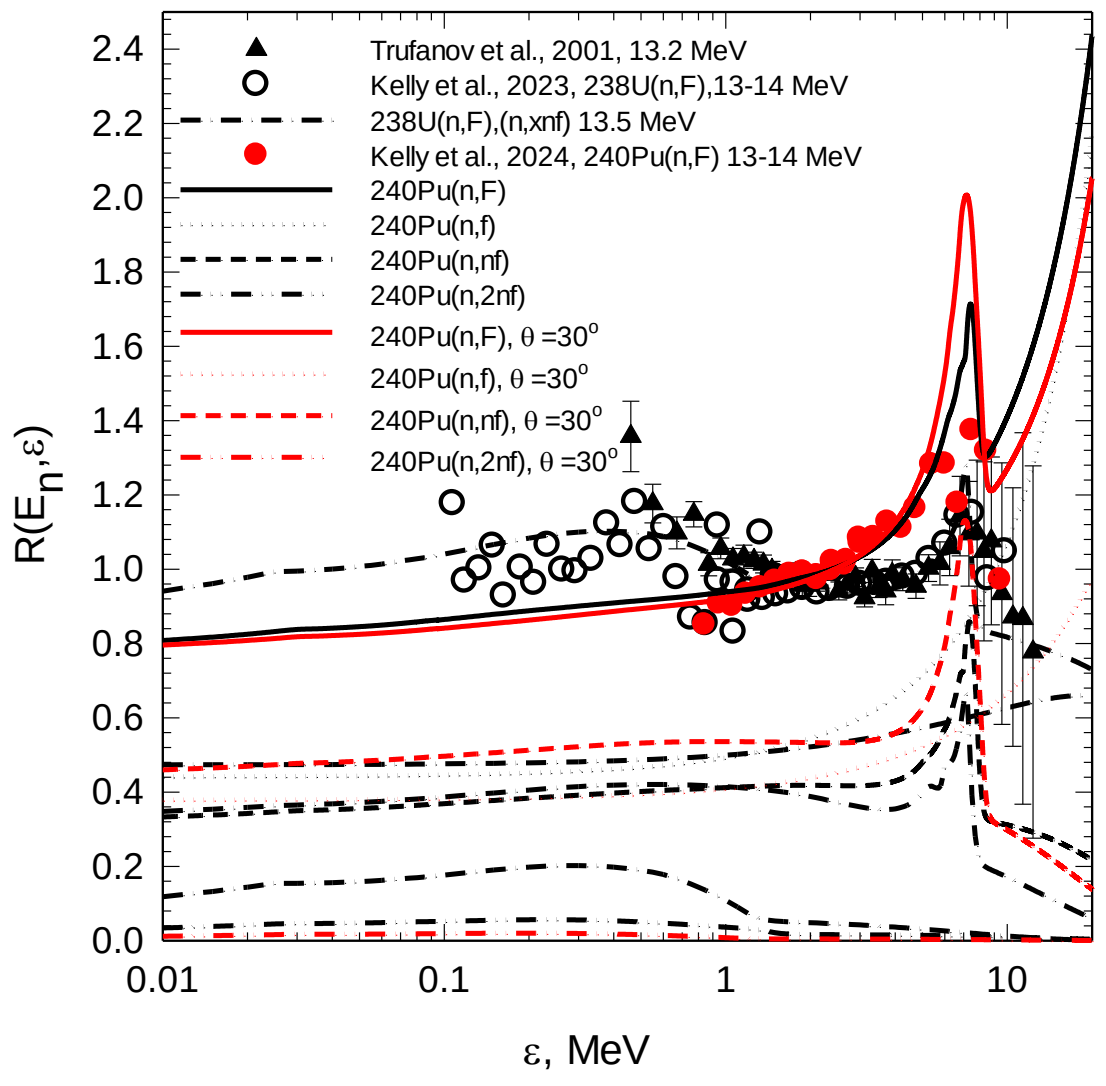

Fig.9. Ratios of partial components of PFNS of $^{240}$Pu($n$, $F$) and $^{238}$U($n$, $F$) at $E_n$ = 13.5 MeV relative to Maxwellian type distribution with $T$ = 1.4241 MeV: • — $^{240}$Pu($n$, $F$) [12]; —— – $^{240}$Pu($n$, $F$); • • • — $^{240}$Pu($n$, $F$); – – – – $^{240}$Pu($n$, $nf$); – –• • – – – $^{240}$Pu($n$, 2$nf$); —— – $^{240}$Pu ($n$, $F$), $\theta$ =30°. • • • – $^{240}$Pu($n$, $F$), $\theta$=30°; – – – – $^{240}$Pu($n$, $nf$) ), $\theta$=30°; – –• • – – – $^{240}$Pu($n$, 2$nf$) ), $\theta$=30°; ○ — $^{238}$U($n$, $F$) [8]; ▲— $^{238}$U($n$, $F$) [54]; – • – – $^{238}$U($n$, $F$), $^{238}$U($n$, $nf$).

Measurements [12, 52] encompass narrower energy range of prompt fission neutrons, than do measurements of PFNS with monochromatic neutron beams [54–56]. That means they don't envision the step-like structure due to $^{238}$U($n$, 2$nf$)$^1$ neutrons [54–56]. Figures 10 and 11 show the partial contributions of $^{240}$Pu($n$, $f$), $^{240}$Pu($n$, $nf$) and $^{240}$Pu($n$, 2$nf$) contributions to the $^{240}$Pu($n$, $F$) PFNS at $E_n$ ~17.7 MeV at $\theta$ ~90° and $\theta$ ~30°. The step-like structure in PFNS due to $^{240}$Pu($n$, 2$nf$)$^1$ neutrons is not pronounced neither at $\theta$ ~90° or $\theta$ ~30°. Partial contributions of $^{238}$U($n$, $f$), $^{238}$U($n$, $nf$) and $^{238}$U($n$, 2$nf$) contributions to $^{238}$U($n$, $F$) PFNS at $E_n$ ~17.7 MeV at $\theta$ ~90° are much different, than those of $^{240}$Pu($n$, $F$) reaction. The hardening of PFNS $\langle E \rangle$ at $\theta$ ~30° is due to the increase of the second chance fission contribution and hardening of exclusive pre-fission neutron spectra $\frac{d^2\sigma^1_{nnf}(\varepsilon, E_n, \theta)}{d\varepsilon d\theta}$.

Angular distribution of $^{239}$Pu(*n,xnf*) pre-fission neutrons at $E_n$ ~14–18 MeV, retrieved in [17] from $^{239}$Pu(*n, F*) PFNS, was interpreted in [1, 18]. Estimate of pre-fission neutrons contribution in [17] obtained as a difference of observed PFNS and some simple estimate of post-fission neutrons evaporated from fission fragments of $^{239}$Pu(*n,xnf*) reactions. Though the procedure adopted in [17] is susceptible to systematic uncertainties, since post-fission neutrons emerge from any of $^{239}$Pu(*n,xnf*) reaction fission fragments, it seems hidden normalizations were used in [17]. It is possible that the normalization was accomplished in the energy range $\varepsilon>E_{nnf1}$, where only neutrons emitted from the fission fragments are observed. Figures 1 and 2 show the influence of forward and backward neutron emission on PFNS shape and average PFNS energies $\langle E\rangle$ at $E_n \gtrsim$ 12 MeV.

Fig. 1 shows measured $^{239}$Pu(*n, F*) PFNS signature $R^{\exp}(\varepsilon, E_n, \Delta\theta, \Delta\theta^1)$ and calculated ratios of $R(\varepsilon, E_n, \Delta\theta, \Delta\theta^1)$ for $^{240}$Pu(*n, F*) and $^{239}$Pu(*n, F*), lumped contributions of $E_n$~15–17.5 MeV and $\Delta\theta$ ~35°–40° (forward direction) and $\Delta\theta^1$=130°–140° (backward direction)

$$R(\varepsilon,15\div17.5)\approx\frac{\int_{15}^{17.5}\nu_p(E_n,\approx30^o)\sigma_{nF}(E_n,\approx30^o)S(\varepsilon,E_n,\theta\approx30^o)\varphi(E_n)dE_n}{\int_{15}^{17.5}\nu_p(E_n,\theta\approx135^o)\sigma_{nF}(E_n,\theta\approx135^o)S(\varepsilon,E_n,\theta\approx135^o)\varphi(E_n)dE_n},\qquad(20)$$

here $\varphi(E_n)$ is the incident neutron spectrum, which is unknown. Spectra $S(\varepsilon, E_n, \theta)$ normalized to unity. As the first order approximation $R(\varepsilon,15\div17.5)$ calculated as a ratio of $\nu_p(E_n,\theta)\sigma_{nF}(E_n,\theta)S(\varepsilon,E_n\approx15-17.5,\Delta\theta)/\nu_p(E_n,\theta)\sigma_{nF}(E_n,\theta)S(\varepsilon,E_n\approx15-17.5,\Delta\theta^1)$ for $E_n$ ~ 15 MeV, $E_n$ ~ 16 MeV, $E_n$ ~ 17 MeV and $E_n$ ~ 17.5 MeV. Values of $\nu_p(E_n,\theta)$ and $\sigma_{nF}(E_n,\theta)$ for $^{240}$Pu(*n, F*) and $^{239}$Pu(*n, F*) were calculated at the same energies $E_n$, as those in $S(\varepsilon,E_n\approx15-17.5,\Delta\theta)$ or $R(\varepsilon, E_n, \Delta\theta, \Delta\theta^1)$. In case of angular dependent observables for $^{240}$Pu(*n, F*) hidden structures in lumped $R(\varepsilon,15\div17.5)$ constituents (for monochromatic beams) are smoothed. $R^{\exp}(\varepsilon, E_n, \Delta\theta, \Delta\theta^1)$ and $R(\varepsilon,15\div17.5)$ seem to have similar shapes, but the latter is shifted downwards. Solid line of $R(\varepsilon,15\div17.5)$ at Fig. 1 obtained by assuming in Eq. (20) equality of numerator and denominator values at $\varepsilon$~3–5 MeV energy range, i.e., number of neutrons emitted in forward and backward directions, as adopted in [17]. In case of $^{240}$Pu(*n, F*) and $^{239}$Pu(*n, F*) at $\varepsilon>E_{nnf1}$, both $R^{\exp}(\varepsilon, E_n, \Delta\theta, \Delta\theta^1)$ and $R(\varepsilon,15\div17.5)$ values are less than unity, that might be due to influence of angular dependence of $(n, xnf)^{1,2,3}$ neutron emission on the fission chances distribution. The renormalized ratio $R(\varepsilon, E_n, \Delta\theta, \Delta\theta^1)$ of $^{239}$Pu(*n, F*) seems to be consistent with reconstructed data [17], shown on Fig. 1, while those of $^{240}$Pu(*n, F*) reaction are appreciably higher. Angular dependence of the first pre-fission neutron in reactions $(n, nf)^1$ and $(n, 2nf)^1$ [1, 18] helps to interpret the experimental data trend in case of ratio of average energies for “forward” and “backward” emission of pre-fission neutrons in $^{239}$Pu$(n,xnf)^{1,2,3}$ reaction [17]. The ratio of $\langle E(\theta)\rangle/\langle E(\theta^1)\rangle$ in case of $^{240}$Pu(*n, F*) for “forward”, $\Delta\theta$ ~35°–40°, and “backward”, $\Delta\theta^1$=130°–140° emission of pre-fission neutrons also steeply increases starting from $E_n$~10–12 MeV. However, for $^{240}$Pu($n$, $F$) PFNS the ratio of $\langle E(\theta)\rangle/\langle E(\theta^1)\rangle$ only slightly higher than in case of $^{239}$Pu($n$, $F$). At $E_n$~16 MeV the ratio of $\langle E(\theta)\rangle/\langle E(\theta^1)\rangle$, calculated for $\varepsilon$~0.8–10 MeV energy range abruptly drops. Hard pre-fission $^{240}$Pu($n$,$nf$)$^1$ neutrons are responsible for that drop. Average energies calculated at energy range of $\varepsilon$~1–3 MeV, ratios $\langle E(\theta)\rangle/\langle E(\theta^1)\rangle$ are virtually independent on $E_n$.

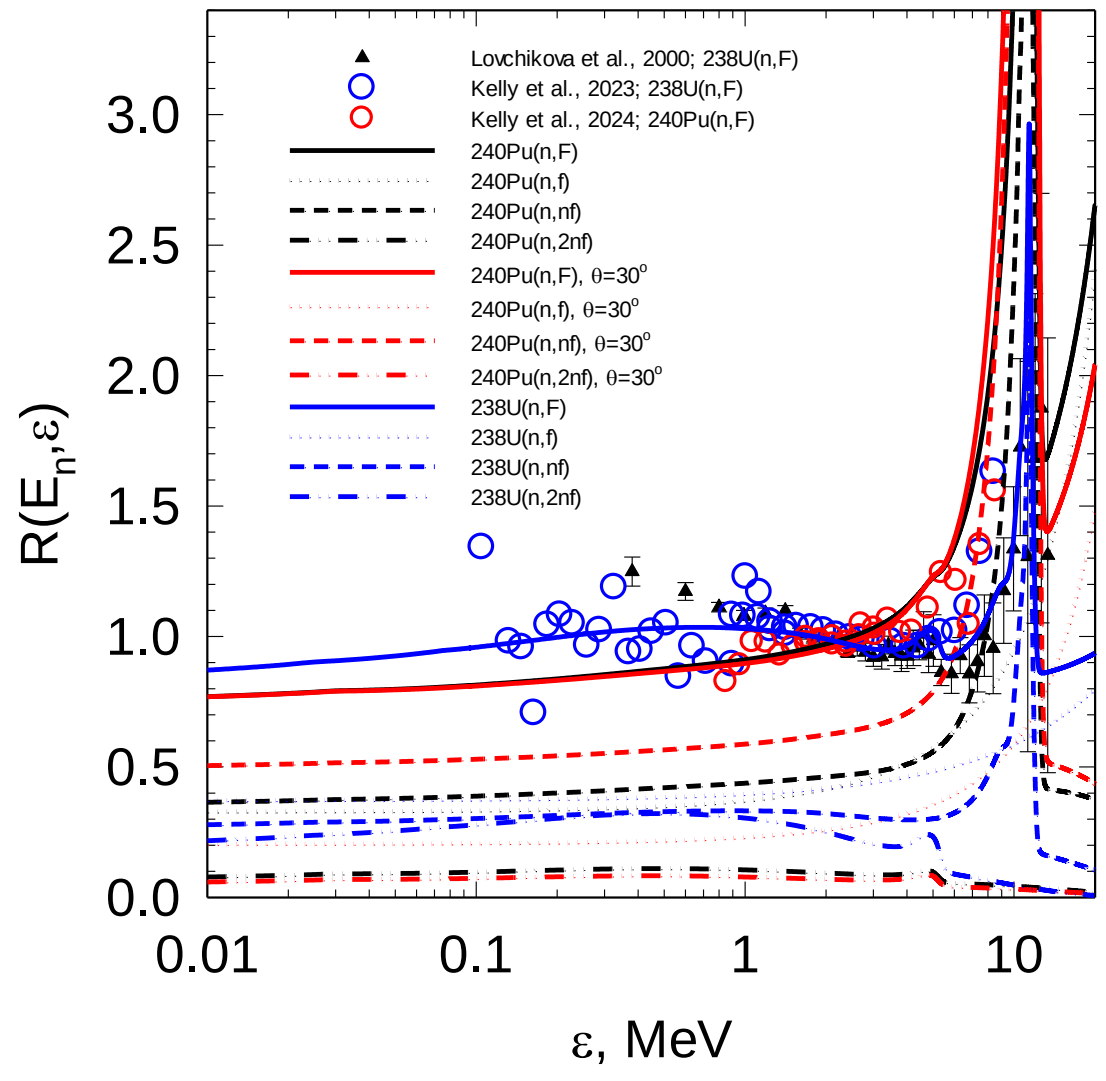

Fig.10. Ratios of partial components of PFNS of $^{240}$Pu($n$, $F$) and $^{238}$U($n$, $F$) at $E_n$ = 17.7 MeV relative to Maxwellian type distribution with $T$ = 1.391867 MeV: —— – $^{240}$Pu($n$, $F$); • • • — $^{240}$Pu($n$, $F$); – – – – $^{240}$Pu($n$, $nf$); – –• • – – – $^{240}$Pu($n$, 2$nf$); —— – $^{240}$Pu ($n$, $F$), $\theta$=30°; • • • – $^{240}$Pu($n$, $F$), $\theta$=30°; – – – – $^{240}$Pu($n$, $nf$) ), $\theta$=30°; – –• • – – – $^{240}$Pu($n$, 2$nf$) ), $\theta$=30°; ○ — $^{238}$U($n$, $F$) [8]; ▲— $^{238}$U($n$, $F$) [54–56];– • – – $^{238}$U($n$, $F$), $^{238}$U($n$, $nf$); ○ — $^{240}$Pu($n$, $F$) [12].

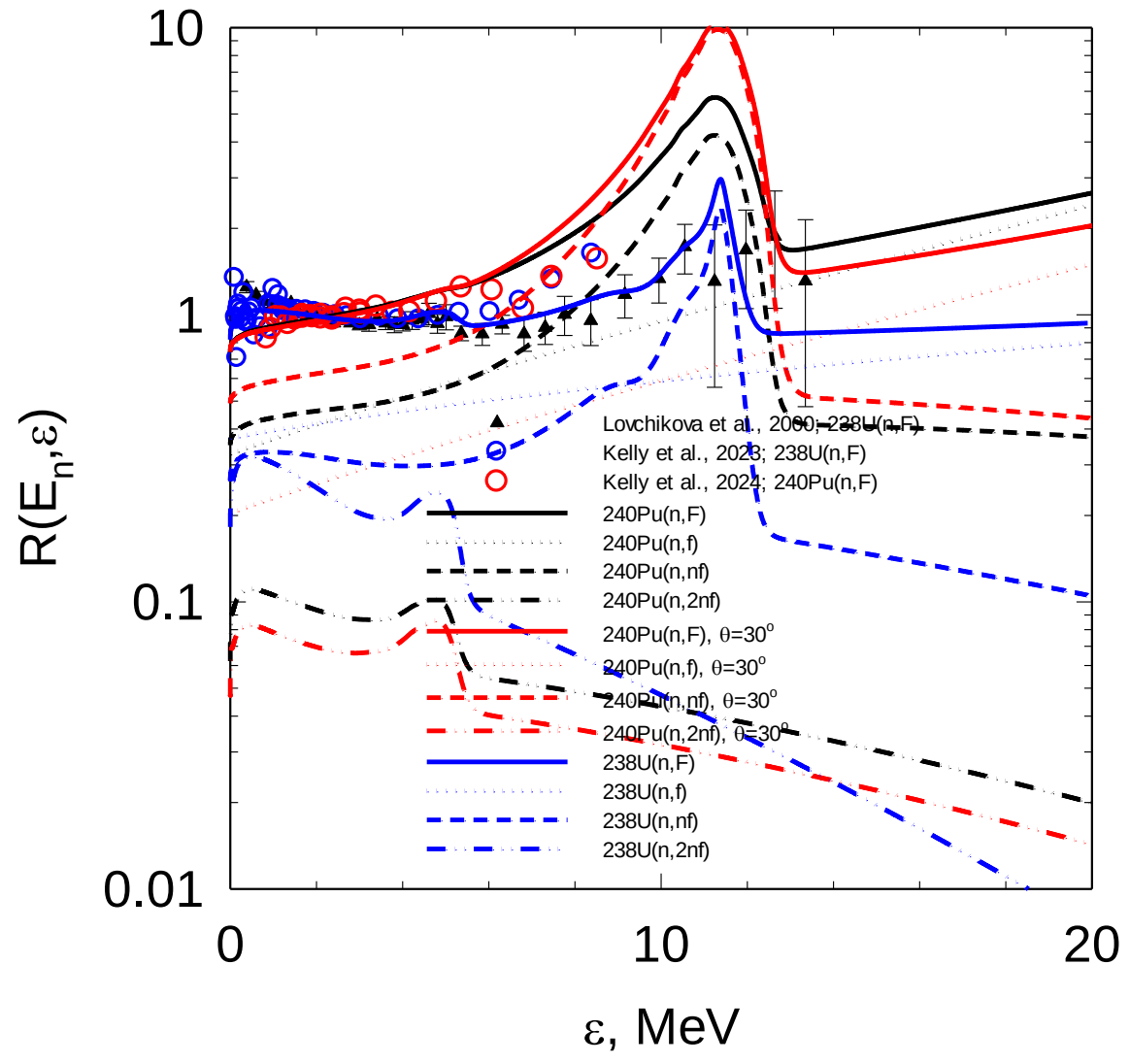

Fig.11. Ratios of partial components of PFNS of $^{240}$Pu($n$, $F$) at $E_n$ = 17.7 MeV relative to Maxwellian type distribution with $T$ = 1.391867 MeV: —— – $^{240}$Pu($n$, $F$); • • • — $^{240}$Pu($n$, $F$); – – – – $^{240}$Pu($n$, $nf$); – –• • – – – $^{240}$Pu($n$, 2$nf$); —— – $^{240}$Pu ($n$, $F$), $\theta$=30°; • • • – $^{240}$Pu($n$, $F$), $\theta$=30°; – – – – $^{240}$Pu($n$, $nf$) ), $\theta$=30°; – –• • – – – $^{240}$Pu($n$, 2$nf$) ), $\theta$=30°; ○ — $^{238}$U($n$, $F$) [8]; ▲— $^{238}$U($n$, $F$) [54–56];– • – – $^{238}$U($n$, $F$), $^{238}$U($n$, $nf$); ○ — $^{240}$Pu($n$, $F$) [12].

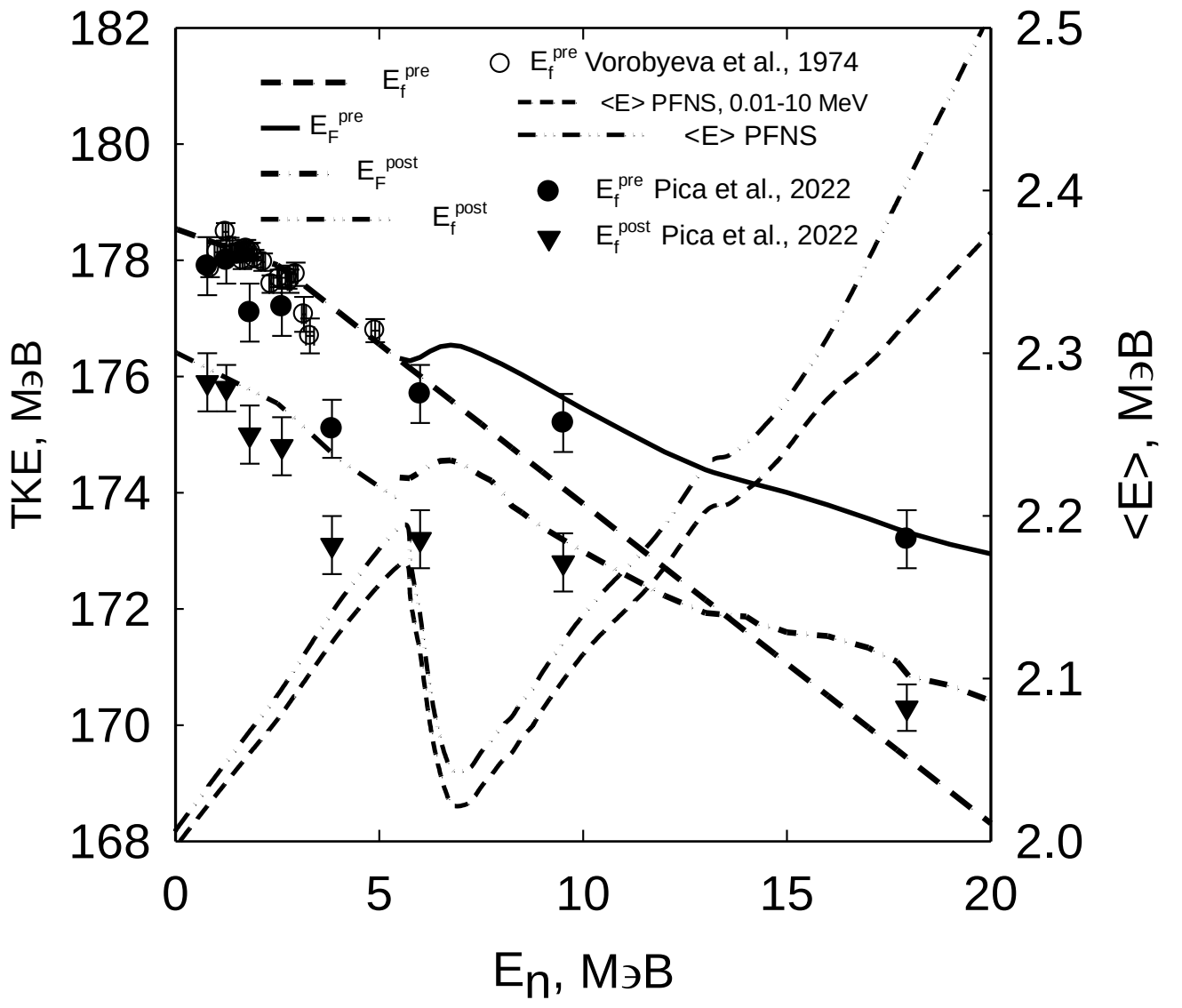

Fig. 12. Total kinetic energy TKE: $^{240}$Pu($n$, $F$), $E_F^{\mathrm{pre}}$ – ——; $^{240}$Pu($n$, $f$), $E_f^{\mathrm{pre}}$ ——— —; $^{240}$Pu ($n$, $F$), $E_F^{\mathrm{post}}$ — —•—; ○ — $^{240}$Pu($n$, $F$), $E_F^{\mathrm{pre}}$ [57]; ● – $^{240}$Pu($n$, $F$) $E_F^{\mathrm{pre}}$ [58]; ▼— $^{240}$Pu($n$, $F$) $E_F^{\mathrm{post}}$ [58 ]; —— , ——— — $\langle E \rangle$ $^{240}$Pu($n$, $F$) in the range ε~0—20 MeV and ε~0.01—10 MeV, respectively.

The ratio of average energies of exclusive neutron spectra of $^{240}$Pu($n$,$nf$)$^1$, $\frac{d^2\sigma^1_{nnf}(\varepsilon, E_n, \theta \approx 30^o)}{d\varepsilon d\theta}$ and $\frac{d^2\sigma^1_{nnf}(\varepsilon, E_n, \theta \approx 135^o)}{d\varepsilon d\theta}$, $\langle E_{n,xnf}(\theta \approx 30^o)\rangle / \langle E_{n,xnf}(\theta^1 \approx 135^o)\rangle$, is much higher than that of $\langle E(\theta)\rangle / \langle E(\theta^1)\rangle$, however it follows the shape of experimental ratio $\langle E(\theta \approx 30^o)\rangle / \langle E(\theta^1 \approx 135^o)\rangle$ [17]. Angular dependence of the ratio of average energies of exclusive neutron spectra of $^{240}$Pu($n$, $2nf$)$^1$ $\frac{d^2\sigma^1_{n2nf}(\varepsilon, E_n, \theta \approx 30^o)}{d\varepsilon d\theta}$ and $\frac{d^2\sigma^1_{n2nf}(\varepsilon, E_n, \theta \approx 150^o)}{d\varepsilon d\theta}$ is much weaker (see Fig. 2).

## VI. AVERAGE TOTAL KINETIC ENERGY TKE

The excitation energy of residual nuclides, after emission of ($n$, $xnf$) neutrons, decreases by the binding energy of emitted neutron $B_{nx}$ and its average kinetic energy as

$$U_x = E_n + B_n - \sum_{x,1\le k\le x} (< E^k_{nxnf}(\theta) > + B_{nx}). \qquad (21)$$

The excitation energy of fission fragments is

$$E_{nx} = E_r - E^{pre}_{fx} + E_n + B_n - \sum_{x,1\le k\le x} \left( \left\langle E^k_{nxnf}(\theta) \right\rangle + B_{nx} \right). \qquad (22)$$

Value of TKE, kinetic energy of fission fragments prior prompt neutron emission (see Fig. 12), $E_F^{pre}$, is approximated by a superposition of partial TKE of $^{241\text{-}x}$Pu nuclides as

$$E_F^{pre}(E_n) = \sum_{x=0}^{X} E_{fx}^{pre}(E_{nx}) \cdot \sigma_{n,xnf} / \sigma_{n,F} \quad . \qquad (23)$$

Kinetic energy of fission fragments, i.e. post-fission fragments after neutron emission (see Fig. 12), $E_F^{post}$, are defined as

$$E_F^{post} \approx E_F^{pre}\left(1 - \nu_{post} / \left(A + 1 - \nu_{pre}\right)\right). \qquad (24)$$

Similar approach was applied in case of $^{239}$Pu(*n, F*) reaction in [1], the structures in TKE [59, 60] are less pronounced than in case of $^{240}$Pu(*n, F*) reaction Similar relation was used for $E_f^{post}$ in [61] at $E_n$<$E_{nnf}$.

## CONCLUSIONS

Analysis of prompt fission neutron spectra of $^{240}$Pu(*n, F*) evidenced correlations of a number of observed data structures with $(n,xnf)^{1...x}$ pre-fission neutrons. Pre-fission neutron spectra turned out to be quite soft as compared with neutrons emitted by excited fission fragments. The net outcome of that is the decrease of $\langle E \rangle$ in the vicinity of the (*n,xnf*) thresholds of $^{240}$Pu(*n, F*). The amplitude of the $\langle E \rangle$ variation is much higher in case of $^{240}$Pu(*n, F*) as compared with $^{239}$Pu(*n, F*). The correlation of PFNS shape with angles of emission of $(n,xnf)^1$ neutrons and emissive fission contributions for $^{240}$Pu(*n, F*) reaction cross section is established. The angular anisotropy of exclusive pre-fission neutron spectra strongly influences the PFNS shapes and their average energies $\langle E \rangle$. These peculiarities are due to varying emissive fission (*n,xnf*) contributions in $^{239}$Pu(*n, F*) and $^{240}$Pu (*n, F*) reactions. Calculated ratio of $\langle E \rangle$ for "forward" and "backward" emission of pre-fission neutrons steeply rises with the increase of average energies of exclusive pre-fission neutron spectra $^{240}$Pu$(n,xnf)^{1...x}$.

The calculated anisotropy of pre-fission neutrons of $^{240}$Pu(*n,xnf*) reaction is a bit higher than in case of $^{239}$Pu(*n, F*). That might be due to correlation of anisotropy of pre-fission neutrons with contribution of emissive fission (*n, nf*) to the observed fission cross section, PFNS and angular anisotropy of NES. In case of $^{240}$Pu(*n, F*) and $^{239}$Pu(*n, F*) at $\varepsilon > E_{nnf1}$, both $R^{\exp}$ and $R(\varepsilon, 15-17.5)$ are less then unity, that also might be due to influence of angular dependence of (*n,xnf*) neutron emission on the fission chances distribution.

The structures in $E_F^{pre}$ and $E_F^{post}$ are shown to be correlated with emission of $^{240}$Pu (*n, xnf*) neutrons due to cooling of the fissionning nuclides/.